\documentclass[numberedappendix]{emulateapj}
\usepackage{natbib,apjfonts,lscape}
\newcommand{\lya}{Ly$\alpha$}

\newcommand{\kms}{\,km\,s$^{-1}$}

\newcommand{\KPC}{{\rm \,kpc}}

\newcommand{\eg}{e.g.,}
\newcommand{\ie}{i.e.,}
\newcommand{\civa}{C{\scriptsize ~IV}~$\lambda 1548$}
\newcommand{\civb}{C{\scriptsize ~IV}~$\lambda 1550$}
\newcommand{\civ}{C{\scriptsize ~IV}}

\newcommand{\siiva}{Si{\scriptsize ~IV}~$\lambda 1393$}

\newcommand{\ra}[3]{#1$^{\rm h}$\,#2$^{\rm m}$\,#3$^{\rm s}$}
\newcommand{\dec}[3]{$#1^\circ\,#2'\,#3''$}

\newcommand{\phl}{\phantom{$<$}}
\newcommand{\phle}{\phantom{$\leq$}}

\journalinfo{Accepted for publication in the Astrophysical Journal}
\slugcomment{To appear in the Astrophysical Journal in November 2007}

\shorttitle{The sightline to Q2343--BX415}
\shortauthors{Rix et al.}

\begin{document}

\title{The sightline to Q2343--BX415: Clues to galaxy formation in a quasar environment\altaffilmark{1}}

\author{\sc Samantha A. Rix\altaffilmark{2},
Max Pettini \altaffilmark{3},
Charles C. Steidel\altaffilmark{4}, 
Naveen A. Reddy\altaffilmark{4,5}, 
Kurt L. Adelberger\altaffilmark{6},
Dawn K. Erb\altaffilmark{7},  and
Alice E. Shapley\altaffilmark{8}}

\altaffiltext{1}{Based on data obtained at the 
W.M. Keck Observatory, which is operated as a scientific partnership 
among the California Institute of Technology, the
University of California, and NASA, and was made possible by the 
generous financial support of the W.M. Keck Foundation.}
\altaffiltext{2}{Isaac Newton Group of Telescopes, Apartado de Correos 321, E-38700 Santa Cruz de La Palma, Canary Islands, Spain}
\altaffiltext{3}{Institute of Astronomy, Madingley Road, Cambridge, CB3 0HA, UK}
\altaffiltext{4}{Astronomy Option, California Institute of Technology, MS 105-24, 
Pasadena, CA 91125}
\altaffiltext{5}{National Optical Astronomy Observatory, 950 North Cherry Street, 
Tucson, AZ 85719}
\altaffiltext{6}{McKinsey and Company, 1420 Fifth Avenue, Suite 3100, Seattle, WA 98101}
\altaffiltext{7}{Harvard-Smithsonian Center for Astrophysics, 60 Garden Street, Cambridge, MA 02138}
\altaffiltext{8}{Department of Astrophysical Sciences, Peyton Hall-Ivy Lane, Princeton, NJ 08544}

\begin{abstract}
We have discovered a strong damped Lyman alpha system essentially
coincident in redshift with the faint QSO Q2343--BX415 (${\cal R} =
20.2$, $z_{\rm em} = 2.57393$).  Such `proximate' damped systems
(PDLAs) are rare and deserving of further study as potential probes of
the environments where AGN are found.  Follow-up observations of the
spectrum of Q2343--BX415 at intermediate spectral resolution reveal
that the metal lines associated with the PDLA consist of two sets of
absorption components. One set of components is apparently moving
towards the quasar with velocities of $\sim 150 - 600$\kms; this gas
is highly ionized and does not fully cover the continuum source,
suggesting that it is physically close to the active nucleus of the
galaxy. Most of the neutral gas in the PDLA is blueshifted relative to
the QSO by $\sim 160$\kms.  We explore the possibility that the PDLA
arises in the outflowing interstellar medium of the host galaxy of
Q2343--BX415; we conclude that this interpretation is supported by the
presence of strong C\,{\sc iv} and N\,{\sc v} absorption at nearby
velocities (the latter in particular requiring a non-thermal ionizing
source), and by the intense radiation field longward of the Lyman
limit implied by the high C\,{\sc ii}$^{\ast}$/H\,{\sc i} ratio. If
Q2343--BX415 is the main source of these UV photons, then the PDLA is
located at either $\sim 8$ or $\sim 37$\,kpc from the active
nucleus. Alternatively, the absorber may be a foreground star-forming
galaxy unrelated to the quasar and coincidentally at the same
redshift, but our deep imaging and follow-up spectroscopy of the field
of Q2343--BX415 has not yet produced a likely candidate.  We measure
the abundances of 14 elements in the PDLA, finding an overall
metallicity of $\sim 1/5$ solar and a normal pattern of relative
element abundances for this metallicity, including moderate depletions
of refractory elements and no overabundance of the $\alpha$-capture
elements relative to the Fe group (once depletions have been taken
into account). Thus, in this PDLA there is no evidence for the
super-solar metallicities that have been claimed for some proximate,
high ionization, systems.  In addition to the PDLA, our spectrum of
Q2343--BX415 shows absorption by 11 intervening systems at $z_{\rm
abs} < z_{\rm em}$; we find possible galaxy identifications for four
of them from an as yet incomplete spectroscopic survey of galaxies
within $\sim 1$\,arcmin of the QSO sight-line.
\end{abstract}


\section{INTRODUCTION}
Absorption line systems at redshifts close to the emission redshift 
of the QSO against which they are viewed have traditionally
been considered to be somewhat different from the bulk of the truly 
intervening absorbers and, for this reason, are normally excluded
from statistical studies of QSO absorption spectra. Possible differences
of these so-called `proximate' systems compared to `normal' ones 
may include: (i)~they could be produced by 
material ejected by the quasar or in the host galaxy interstellar
medium;  (ii)~their degree
of ionization may be influenced by the proximity of the 
quasar's radiation field \citep[\eg][]{foltz86};
and (iii)~environmental effects could come into play if the
quasars are located in overdense regions 
\citep[\eg][]{sargent82}. 
Such reservations also apply to 
damped Lyman alpha
systems, the absorbers at the upper end
of the distribution of neutral hydrogen 
column densities with
$N$(H\,{\sc i})\,$ \geq 2 \times 10^{20}$\,cm$^{-2}$
\citep{wolfe05}, if located close to the QSO redshift.
Indeed, a recent study by \cite{russell06} has
found that the number density of DLAs almost doubles in the
vicinity of QSOs (within 6000\,\kms\ of the emission
redshift), leading those authors to speculate that the excess 
may result from galaxies in the same cluster or supercluster as the
quasar. 

\begin{figure*}[t]
\begin{center}
\centerline{\includegraphics[angle=270,width=0.975\textwidth]{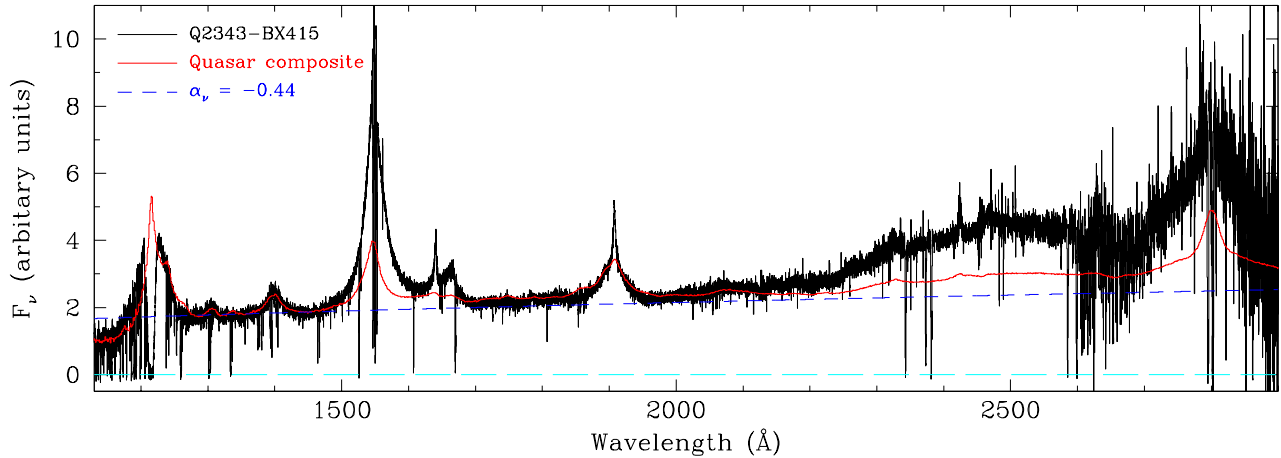}}
\caption{Spectrum of Q2343--BX415 ({\em histogram}) in the systemic
rest-frame of the quasar ($z_{\rm sys}=2.57393$). Overplotted is the
median composite quasar spectrum from the SDSS \citep[{\em continuous
line}; ][]{vandenberk01}, along with its best fitting power law slope
$\alpha_\nu = -0.44$ ({\em short-dashed line}). The composite spectrum was
multiplicatively scaled to compare its continuum slope with that of
Q2343--BX415.}\label{fig:plot_bx415_slope}
\end{center}
\end{figure*}

While it may therefore be justified to exclude 
proximate damped systems, or PDLAs, when
considering the statistical
properties of DLAs as a whole,
the very fact that they may be a different 
population makes them interesting in their own right. 
Can we learn more about the quasar environment
from PDLAs?  Are their chemical
properties different from those of intervening DLAs? 
Supersolar metal abundances have been 
claimed for gas close to quasar nuclei \citep[][]{hamann99,nagao06a}
and for high ionization associated systems \citep[][]{petitjean94},
but such estimates are much more model dependent 
than the relatively straightforward determination 
of element abundances in DLAs
\citep[see the discussion by][]{pettini07}.
Of course there is no reason to assume 
that all PDLAs have the same origin, and therefore similar properties;
indeed some may simply be chance alignments, comparable to the
intervening absorbers, while others could form part of the same
group or cluster as the QSO, 
with perhaps only a small fraction originating from the host
galaxy itself.

Until recently there have been very few studies of PDLAs, partly
because these systems are relatively rare (since DLAs themselves are
rare systems), but also because they have often been ignored.  While
the handful of known cases was boosted slightly with the CORALS survey
by \citet{ellison02}, it is the Sloan Digital Sky Survey (SDSS) that
is offering the greatest opportunity to significantly increase the
sample size \citep[see \eg][]{russell06,prochaska07}. Consequently it
should soon be possible to examine more generally the statistical
properties of the PDLA population.

The few PDLAs studied to date have been found to be broadly 
similar to other DLAs, although there is some evidence of 
enhanced \lya\ emission \citep{pettini95,moller93,moller98}. 
This emission has been
attributed to various causes including (i)~photoionization from the
quasar, (ii)~the reduced line blanketing from the \lya\ forest, due to
the proximity effect, and (iii)~increased star formation in the quasar
environment (either directly through gravitational interaction or
indirectly due to the fact that quasar activity tends to be located near
to galaxies that are actively forming stars).
The similarity with intervening DLAs also appears to extend 
to the chemical properties of the few PDLAs studied so far,
which do not appear to be statistically different from the 
rest of the population of damped absorbers \citep[][]{akerman05,lu96}.
One difficulty with all of these comparisons, however, 
is the uncertainty by up to
several thousand \kms\ in the systemic redshifts of the bright
quasars against which PDLAs have been observed; without 
knowledge of the QSO true redshift, it is not possible to establish
whether a proximate system is associated with the QSO  
environment or is located tens of Mpc in front of it.

\begin{deluxetable*}{lclccl}
\tablewidth{370pt}
\tablecaption{\textsc{The Systemic Redshift of Q2343--BX415}}
\tablehead{
  \colhead{Transition}
& \colhead{$\lambda_{\rm vac}$ (\AA)\tablenotemark{a}}
& \colhead{Instrument}
& \colhead{$\lambda_{\rm obs}$ (\AA)\tablenotemark{b}}
& \colhead{FWHM (\kms)\tablenotemark{c}}
& \colhead{$z_{\rm sys}$\tablenotemark{d}}
}
\startdata
\\
~He {\scriptsize  II} $\lambda 1640$\tablenotemark{e}              & 1640.418\phm{:} & ESI     & \phn5862.9 $\pm$ 0.1 & \phn612  $\pm$ \phn19    & 2.57403\phm{:} $\pm$ 0.00008\phn\\
~C {\scriptsize  III}] $\lambda\lambda 1907,1909$\tablenotemark{e} & 1907.503:       & ESI     & \phn6817.8 $\pm$ 0.1 & \phn493  $\pm$ \phn25    & 2.57419       $\pm$ 0.00007\tablenotemark{f}\\
~H$\beta$                                                          & 4862.683\phm{:} & NIRSPEC & 17380.6    $\pm$ 4.3 &    1550 $\pm$       200  & 2.57428\phm{:} $\pm$ 0.00088\phn\\
~[O {\scriptsize  III}] $\lambda 4959$                             & 4960.295\phm{:} & NIRSPEC & 17728.0    $\pm$ 0.7 & \phn570  $\pm$ \phm{20}8 & 2.57397\phm{:} $\pm$ 0.00014\phn\\
~[O {\scriptsize  III}] $\lambda 5007$                             & 5008.240\phm{:} & NIRSPEC & 17899.1    $\pm$ 0.2 & \phn570  $\pm$ \phm{20}8 & 2.57393\phm{:} $\pm$ 0.00004\phn\\
\enddata
\tablenotetext{a}{Vacuum wavelengths of the transition.}
\tablenotetext{b}{Observed central wavelengths of the Gaussian fits.}
\tablenotetext{c}{FWHM of the emission line after correcting for the instrumental resolutions of either 45.5\kms (ESI) or 214\kms (NIRSPEC).}
\tablenotetext{d}{Systemic redshift.}
\tablenotetext{e}{Both He{\scriptsize\,II}~$\lambda 1640$ and C{\scriptsize\,III}]~$\lambda\lambda 1907,1909$ display broad emission features with a narrow component superposed. For the purposes of this analysis we simultaneously fitted both the broad and narrow components, but only quote the fit to the narrow line region.}
\tablenotetext{f}{Uncertainties are introduced into the determination 
of $z_{\rm sys}$ from C\,{\sc iii}] due to the unknown intensity ratio between the two lines at 1906.683\,\AA\ and 1908.734\,\AA, and hence the unknown value of 
$\lambda_{\rm vac}$ . 
Here we assume the regime of low electron densities, where I(1906.683)/I(1908.734) $\rightarrow$ 1.5. If the electron density were higher, this ratio would decrease and result in a higher value of $\lambda_{\rm vac}$ for the blend, and therefore a lower systemic redshift. 
The FWHM quoted corresponds to the fit over both transitions.\\ \\}
\label{tab:systemic_redshift}
\end{deluxetable*}


\subsection{Q2343--BX415}
In this paper we present observations of a new PDLA
in the spectrum of the faint (${\cal R} = 20.22$)
$z_{\rm em} = 2.574$ QSO Q2343--BX415 at coordinates 
$\alpha$ = \ra{23}{46}{25.4}, $\delta$ = \dec{+12}{47}{44} (J2000).
The QSO and its associated DLA
were discovered in the course of a survey
for galaxies at $z \sim 2$ in the field of the previously
known, brighter  (${\cal R} = 17.0$) quasar  
Q2343+1232 at a similar redshift ($z_{\rm em} = 2.573$);
the aim of the survey is to study the connection
between star-forming galaxies and the absorption
seen in QSO spectra at redshifts 
$z =\simeq 2 - 2.5$ \citep{steidel04, adelberger05}.
The discovery spectrum of Q2343--BX415, obtained with the 
Low Resolution Imaging Spectrograph on 
the Keck\,I telescope, revealed the presence of
a strong damped Ly$\alpha$ line essentially
coincident in redshift with the Ly$\alpha$ emission 
line, raising the possibility that this PDLA may be 
associated with the host galaxy of the QSO. 
Whatever its origin, PDLAs so close in redshift
to the background QSO are very rare and warrant further
study. 

We therefore secured further observations 
of Q2343--BX415 with the Keck\,II telescope, 
in the optical 
with the Echelle Spectrograph and
Imager \citep[ESI; ][]{sheinis02} 
and in the near infrared with the
Near Infrared Spectrograph \citep[NIRSPEC;][]{mclean98},
for the purpose of studying in detail 
the chemical and kinematic properties of the PDLA and
establishing its relationship to the QSO itself. These 
observations are the subject of the present paper,
which is organized as follows.
In \S\ref{sec:observations} we briefly describe the 
ESI observations and the steps in the data reduction 
process. In \S\ref{sec:spectrum_bx415} we give an
overall description of the optical spectrum of Q2343--BX415,
measure its systemic redshift from the near-IR observations,
and derive the column density of the PDLA by
decomposing the \lya\ line into its emission and
absorption components. In \S\ref{sec:kinematics} we discuss
the kinematics of the metal absorption lines associated
with the PDLA, which are found to consist of two sets
of components, one redshifted and the other one
blueshifted relative to $z_{\rm sys}$. Measurements
pertaining to the two sets of components are 
collected in Table~\ref{tab:table_dla_ew}.
\S\ref{sec:abundances} deals with the derivation of 
the abundances of 14 elements in the PDLA, while
\S\ref{sec:partial_coverage} and \S\ref{sec:ionization_corrections}
examine critically how these abundance determinations may be
compromised by, respectively,  partial coverage of the 
QSO continuum by the absorbing gas and the presence of 
(partly) ionized gas. In \S\ref{sec:CII*} we use the ratio
$N$(C\,{\sc ii}$^{\ast}$)/$N$(H\,{\sc i}) to deduce a 
conservative limit to the physical distance between
the PDLA and Q2343--BX415. We bring all of these findings
together in the Discussion of \S\ref{sec:discussion}. 

In addition to the PDLA, our 
ESI spectrum of Q2343--BX415 includes absorption lines
from 11 intervening absorption systems at $z_{\rm abs} < z_{\rm em}$.
We catalog them in Appendix~\ref{sec:intervening_systems},
where we also compare their redshifts to those of galaxies within
$\sim 1$\,arcmin of the line of sight to Q2343--BX415.
Throughout the paper we adopt 
a cosmology with $\Omega_{\rm M} = 0.3$,
$\Omega_{\Lambda} = 0.7$, and
$H_0 = 70$\kms\ Mpc$^{-1}$.

\section{OBSERVATIONS AND DATA REDUCTION}\label{sec:observations}

ESI is a highly suitable instrument for 
observations of absorption systems in faint  QSOs,
offering a combination of high efficiency,
wide wavelength coverage (from 
4000 to 10500\,\AA\ spread over
ten spectral orders) and a moderately
high spectral dispersion of 11.5\,\kms\ per
15\,$\mu$m detector pixel.
We observed Q2343--BX415 on the nights of
4 and 5 September 2002 
for a total exposure time of 15,600\,s
(four 3600\,s + one 1200\,s exposures).
The atmospheric conditions were mostly clear
with seeing of $\sim 0.6$\,arcsec FWHM;
the spectrograph slit width was set at $0.75$\,arcsec
projecting to $\sim 4$ CCD pixels.

The individual ESI exposures were 
reduced with purpose-written IRAF tasks,
which implement the conventional steps of
bias subtraction, flat-fielding, cosmic-ray rejection
and background subtraction prior to
co-adding the individual one-dimensional
spectra.
Wavelength calibration used the reference
spectra of Cu-Ar and Hg-Xe-Ne lamps, 
and an absolute flux scale was provided
by observations of the standard star
G191B2B \citep{oke90}. 
The intrinsically smooth spectrum
of this white dwarf also allowed us to correct for
telluric absorption at red wavelengths.\footnote{
We did not apply this correction at wavelengths
longward of 9250\,\AA\ where S/N\,$< 20$.}
Finally, the QSO spectrum was mapped onto
a vacuum heliocentric wavelength scale.
For the purpose of studying absorption lines,
we fitted the QSO intrinsic spectrum with 
a spline curve, paying special attention 
to the fit around the strong emission lines on which
resonance absorption lines from the same transition
are superposed; division by this spline fit then 
provided a normalized spectrum suitable for
absorption line measurements. 

The signal-to-noise ratio of the data varies from
S/N\,$\sim 20$ to $\sim 170$ per 0.25\,\AA\ wavelength 
bin over the interval 5000--9250\,\AA, being 
higher in the QSO broad emission lines
and in the center of spectral orders, and lower
in the continuum and at the edges of orders.
It falls to S/N\,$\sim 10$ by 4250\,\AA\ in the blue
and 10,000\,\AA\ in the red.
The spectral resolution is FWHM\,$= 45.5 \pm 0.3$\,\kms, sampled with
$\sim 4$ CCD pixels.

\begin{figure*}
\vspace*{0.5cm}
\centerline{\includegraphics[width=0.9\textwidth]{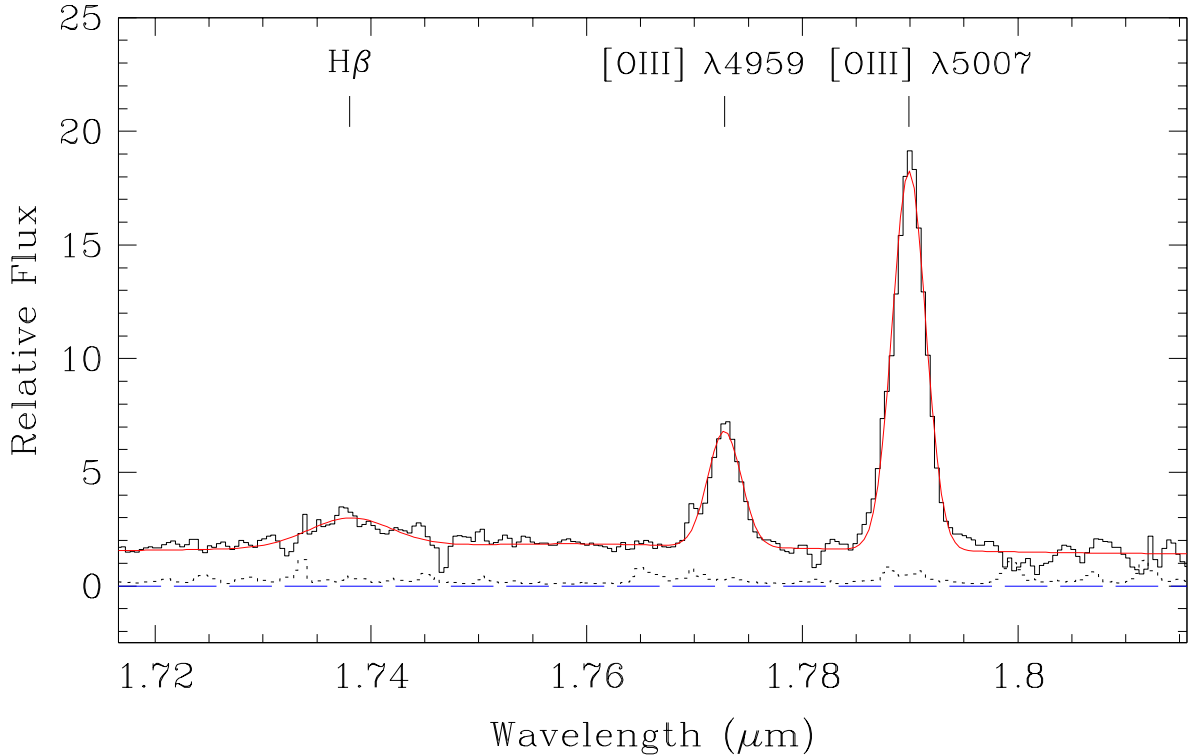}}
\caption{Portion of the NIRSPEC spectrum of 
Q2343--BX415 ({\em histogram})
and its one sigma error spectrum ({\em dotted line}). 
Overplotted are our Gaussian fits to the [O\,{\sc iii}]
and H$\beta$  emission lines ({\em continuous line}), 
superposed on the underlying continuum level.
The three lines give a mean systemic redshift for the QSO of
$\langle z_{\rm sys} \rangle = 2.57393 \pm
0.00005$. }\label{fig:plot_nir_elf}
\vspace{0.5cm}
\end{figure*}

\section{THE SPECTRUM OF Q2343--BX415}\label{sec:spectrum_bx415}

The reduced ESI spectrum of Q2343--BX415 before normalization
is reproduced in Figure\,\ref{fig:plot_bx415_slope}.
In the rest frame of the QSO, at $z_{\rm sys}=2.57393$
(see \S\ref{subsec:systemic_redshift}), our data extend
from $\lambda_0 = 1130$ to 2900\,\AA\ and include all
the familiar ultraviolet emission lines, from 
Ly$\alpha$ to Mg\,{\sc ii}~$\lambda 2798$. 
The strong absorption trough that is
superposed on the \lya\ emission indicates the presence of a
proximate-DLA (see \S\ref{subsec:p-dla}). 
The numerous sharp absorption
lines, which are resolved when the spectrum is plotted at higher
resolution, correspond to metal absorption lines from the PDLA
and from numerous other intervening systems at lower redshifts.

Overplotted on Figure\,\ref{fig:plot_bx415_slope} is 
a portion of the 
template quasar spectrum produced by \cite{vandenberk01}
from the median of over 2200 SDSS QSOs; also
shown is the UV-optical continuum approximated by
a power law with slope $\alpha_\nu = -0.44$.
The broad similarity between the
two spectra suggests that Q2343--BX415 is a fairly typical quasar. 
There are differences too, however. In Q2343--BX415, 
the C\,{\sc iv}~$\lambda 1549$ and Mg\,{\sc ii}~$\lambda 2798$ 
emission lines are stronger, the Fe\,{\sc ii}
broad emission at wavelengths $\lambda \gtrsim 2200$\,\AA\
is also stronger, and there are 
narrow UV emission lines which,
as explained below, help us pin down the systemic redshift
of the QSO.

\subsection{The Systemic Redshift of the Quasar}\label{subsec:systemic_redshift}

As we are interested in investigating the relationship 
between the PDLA and the QSO, it is important to determine 
accurately their respective redshifts. 
It has been known for a number of years that 
in quasars the broad permitted and semi-forbidden
emission lines are not
at the systemic redshift, but are blueshifted 
by up to several thousand \kms\ 
\citep[see \eg][]{gaskell82,espey89}. 
The blueshift is different for different ion stages
and shows a strong correlation with ionization
energy. 

On the other hand, the narrow forbidden lines
and the Balmer lines are closer to the systemic redshift,
with shifts of less than 100\,\kms\ \citep{vandenberk01}.
In order to measure the redshifts
of the [O\,{\sc iii}]~$\lambda \lambda 4959, 5007$
doublet  and H$\beta$, 
we observed Q2343--BX415 in the near-infrared
$H$ band with NIRSPEC on the Keck\,II telescope, 
as part of a larger programme of near-IR
spectroscopy of galaxies and AGN at $z \sim 2$ 
described by \cite{erb06a}. 
Figure~\ref{fig:plot_nir_elf} shows the relevant 
portion of the NIRSPEC spectrum \citep[reduced 
as described by][]{erb06a} together with
Gaussian fits to the three emission lines; the  
parameters of the fits are collected in 
Table~\ref{tab:systemic_redshift}. 
From the weighted average of the three
redshifts we deduce 
$\langle z_{\rm sys} \rangle = 2.57393 \pm 0.00005$
(although the measurement is dominated by the narrow
[O\,{\sc iii}] doublet).
As can be seen from Table~\ref{tab:systemic_redshift},
the redshifts of the narrow components of the ultraviolet
He\,{\sc ii}~$\lambda 1640$ and 
C\,{\sc iii}]~$\lambda \lambda 1907, 1909$
emission lines also agree with this value of $z_{\rm sys}$
within the errors.

We also observed the nearby quasar Q2343+1232 in the $H$ band with
NIRSPEC and measured $ z_{\rm sys} = 2.5727 \pm 0.0004$ from its
[O\,{\sc iii}]\,$\lambda 5007$ line; thus the two QSOs, which are
separated by $\sim 690$\,kpc on the plane of the sky, have redshifts
which differ by only $\sim 100$\kms. The absorption systems in the
optical spectrum of Q2343+1232 have been catalogued by
\citet{sargent88}.  They include a \civ\ system at $z_{\rm abs} =
2.5696$ (which corresponds to a velocity difference $\Delta v =
-365$\,\kms\ from $z_{\rm sys}$ of Q2343--BX415); a DLA at $z_{\rm
abs} = 2.4313$ \citep{lu98}, which differs by $-550$\,\kms\ from the
redshift $z_{\rm abs} = 2.4376$ of one of the intervening \civ\
systems we identify in the spectrum of Q2343--BX415; and a double
\civ\ system at $z_{\rm abs} = 2.1693$, 2.1714, which is also within a
few hundred \kms\ of a rich concentration of galaxies and absorbing
gas near the Q2343--BX415 sight-line. As discussed in
Appendix~\ref{sec:intervening_systems}, these velocity similarities
attest to the large scale structure of gas and galaxies at $z \simeq 2
- 2.6$ seen in this field.

\subsection{The Proximate DLA}\label{subsec:p-dla}

\begin{figure}
\centerline{\includegraphics[width=\columnwidth]{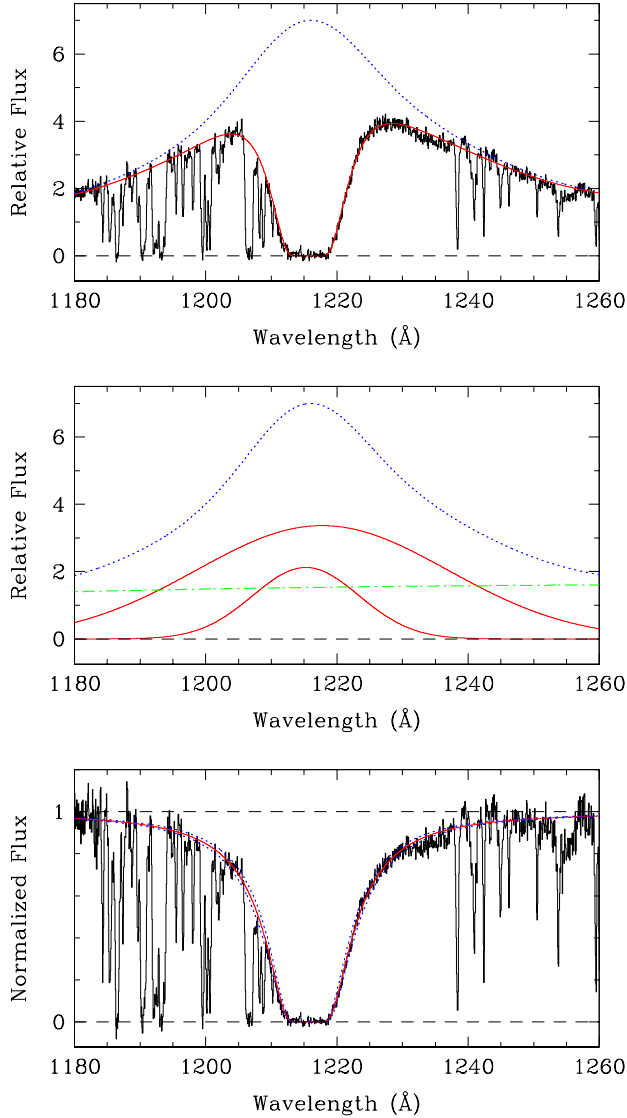}}
\caption{The \lya\ region in Q2343--BX415. {\it Top}: The damped \lya\
absorption trough of the PDLA superposed on the quasar's \lya\
emission line ({\em histogram}). Overplotted is our reconstruction of
the quasar's continuum + \lya\ emission line ({\em dotted line}) and
the PDLA's absorption profile for a neutral hydrogen column density
$N$(H\,{\sc i})\,$=9.5 \times 10^{20}$\,cm$^{-2}$ ({\em continuous
line}). {\it Middle}: The decomposition of the quasar's continuum +
\lya\ emission line ({\em dotted line}) into two Gaussian emission
components ({\em continuous lines}) and the underlying continuum ({\em
dashed-dotted line}). {\it Bottom}: As for the top panel, but normalised
after dividing the observed spectrum by the continuum + \lya\ emission
line fit.  Theoretical damped profiles for column densities
$N$(H\,{\sc i})\,$=(9.5 \pm 1.0) \times 10^{20}$\,cm$^{-2}$ are
overplotted on the data (\textit{continuous and dotted
lines}).\label{fig:plot_HI}}
\end{figure}

In the top panel of Figure~\ref{fig:plot_HI} we have reproduced on an
expanded scale the portion of Figure\,\ref{fig:plot_bx415_slope} that
encompasses the \lya\ region, so that the blend of emission and damped
absorption can be seen clearly.  The decomposition of the profile into
its components is made easier by the high S/N of the ESI data and the
requirement that, once the emission line is divided out, the remaining
absorption should exhibit the characteristic Lorentzian shape of the
damping wings. 

To perform our decomposition, we first subtracted out a fit to the
quasar's underlying continuum level, then used the residual spectrum
for an initial estimate of the shape and intensity of the \lya\
emission line profile. To characterise this emission, we used the {\sc
elf} (Emission Line Fitting) routines in the Starlink {\sc dipso} data
analysis package \citep{howarth03} to fit this estimate with a
(multi-component) Gaussian fit. By dividing the original spectrum by
the sum of the quasar's continuum and this emission line fit, it was
possible to compare the resulting normalised absorption spectrum with
theoretical profiles of damped \lya\ lines of varying H\,{\sc i}
column density, generated by the {\sc dipso} routine {\sc HC}. Any
deviations in the spectra from the theoretical Lorentzian shape were
used to modify the original emission line estimate, and the whole
cycle was repeated. After several iterations we converged onto a
satisfactory solution, shown in the top panel of
Figure~\ref{fig:plot_HI}, for which both damping wings and the core of
the \lya\ line were well fitted by a theoretical
profile. Interestingly, we found that \textit{two} Gaussian
components, separated by $\sim 600$\kms, were required for a
satisfactory fit to the \lya\ emission line\footnote{To accurately
fit the other quasar emission lines observed in the spectrum of
Q2343--BX415, for which the line profiles are much better defined, we
find that two or three Gaussian components are required.} (see middle
panel of Figure~\ref{fig:plot_HI}); other relevant parameters are
collected in Table~\ref{tab:lya_em}.  The normalized absorption
profile is well reproduced by a column density $N$(H\,{\sc i})\,$=9.5
\times 10^{20}$\,cm$^{-2}$ (bottom panel of
Figure~\ref{fig:plot_HI}). We found the column density
derived to be only weakly sensitive to the precise \lya\ emission line
profile adopted (despite the varying quality of the overall fit);
after many trial fits to the composite of emission and absorption, we
estimate the 1$\sigma$ error in $N$(H\,{\sc i}) to be of the order of
10\%, as indicated in the Figure.

\begin{deluxetable}{ccccr}
\tablewidth{175pt}
\tablecaption{\textsc{\lya\ Emission Components in Q2343--BX415}\label{tab:lya_em}}
\tablehead{
  \colhead{Number}
& \colhead{$\lambda_{\rm obs}$}
& \colhead{FWHM}
& \colhead{$z_{\rm em}$}
& \colhead{$\Delta v$\tablenotemark{a}}\\
   \colhead{}
& \colhead{(\AA)}
& \colhead{(\kms)}
& \colhead{}
& \colhead{}
}
\startdata
\\
1 & 4350.45 &    11125      & 2.5786  &   $+390$\\
2 & 4341.61 & \phn4701  & 2.5714  &   $-210$\\
\enddata
\tablenotetext{a}{Velocity difference from $z_{\rm sys} = 2.57393$, determined from
[O\,{\sc iii}] and H$\beta$ (see \S\ref{subsec:systemic_redshift}).\\}
\end{deluxetable}

In Figures~\ref{fig:plot_normalised_vpfit.0} 
through to \ref{fig:plot_normalised_vpfit.6}
we have reproduced the rich absorption spectrum 
of Q2343--BX415 on an expanded scale.
We detect 49 transitions from the PDLA
due to 14 elements ranging, in the periodic table, from
C to Zn and in ionization stages from C\,{\sc i} to 
N\,{\sc v}.  All these absorption lines are
appropriately labelled in 
Figures~\ref{fig:plot_normalised_vpfit.0}--\ref{fig:plot_normalised_vpfit.6}.
We also identify 11 other intervening 
absorption systems, at redshifts lower than that
of the DLA; absorption lines from these systems are
denoted S1 through to S11 and are discussed separately
in the Appendix.

\begin{figure*}
\figurenum{4a}
\includegraphics[width=0.95\textwidth]{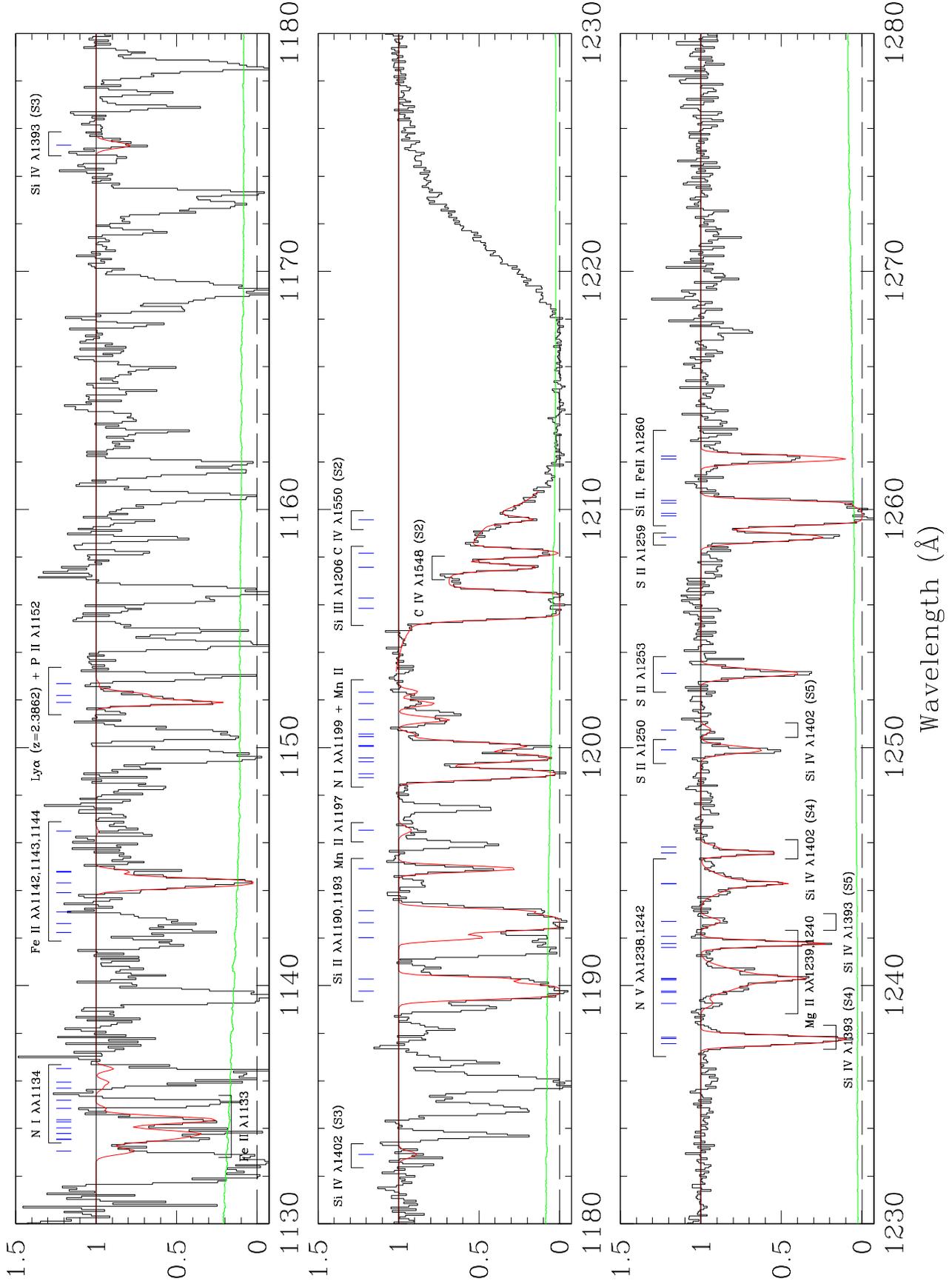}
\caption{Normalised spectrum of Q2343--BX415 ({\em black histogram\/})
in the systemic rest-frame of the quasar 
($z_{\rm sys}=2.57393$). 
The green line near the zero level is the $1 \sigma$ error spectrum.
Absorption line identifications are given for features
originating from (i)~the PDLA and (ii)~any other intervening
absorption line systems along the line of sight; the latter are
labelled by the system number given in Table~\ref{tab:intervening_ews}
(\eg\ S4 and S5 in the bottom panel). 
Overplotted on the spectrum are the absorption line fits
from our VPFIT analysis ({\em red lines\/}), along with the individual
absorption components used in the VPFIT fitting ({\em blue
tickmarks\/}). Blends of lines from different systems were fitted
simultaneously.\label{fig:plot_normalised_vpfit.0}}
\end{figure*}

\begin{figure*}
\figurenum{4b}
\includegraphics[width=0.95\textwidth]{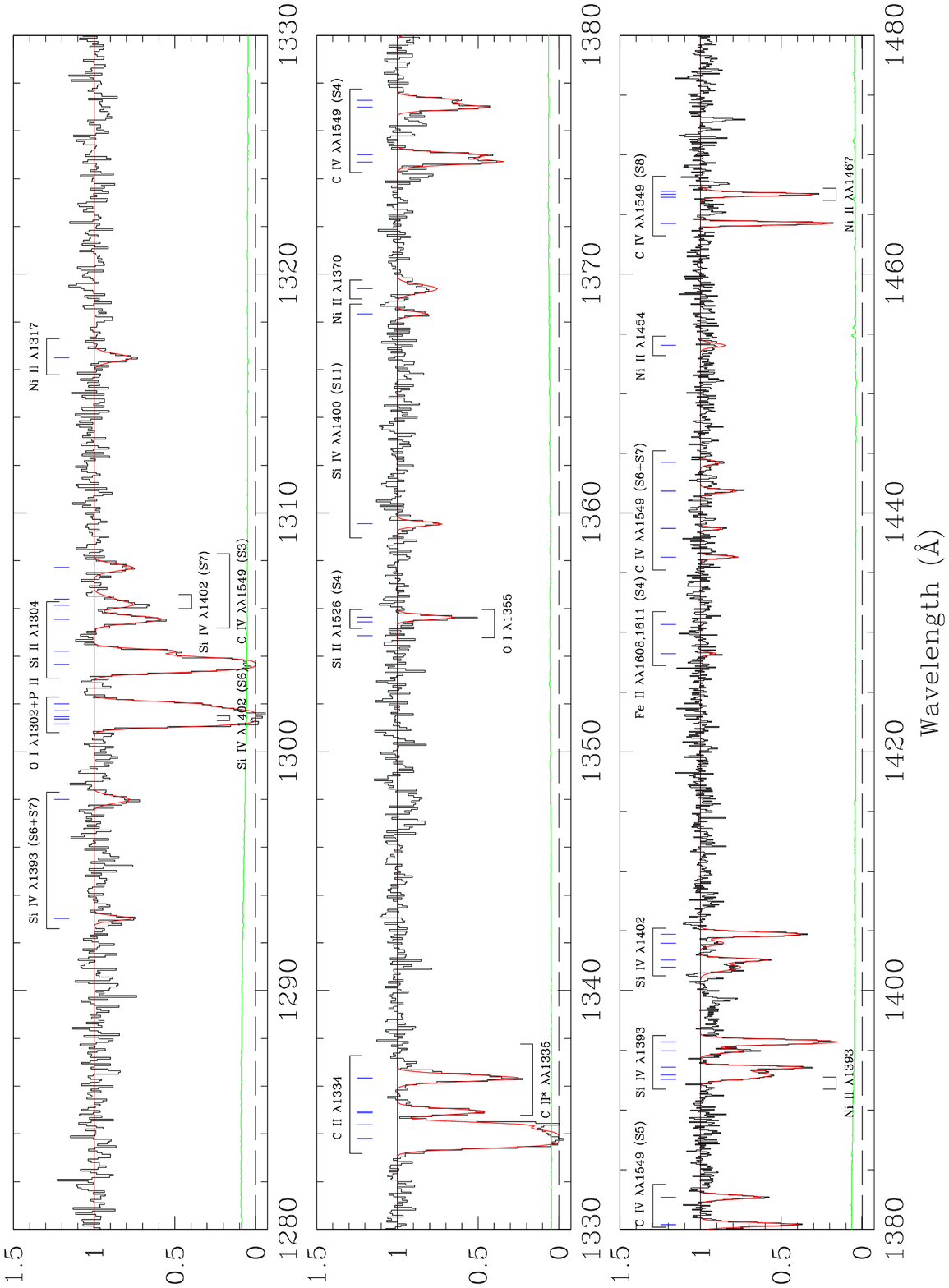}
\caption{continued \label{fig:plot_normalised_vpfit.1}}
\end{figure*}

\begin{figure*}
\figurenum{4c}
\includegraphics[width=0.95\textwidth]{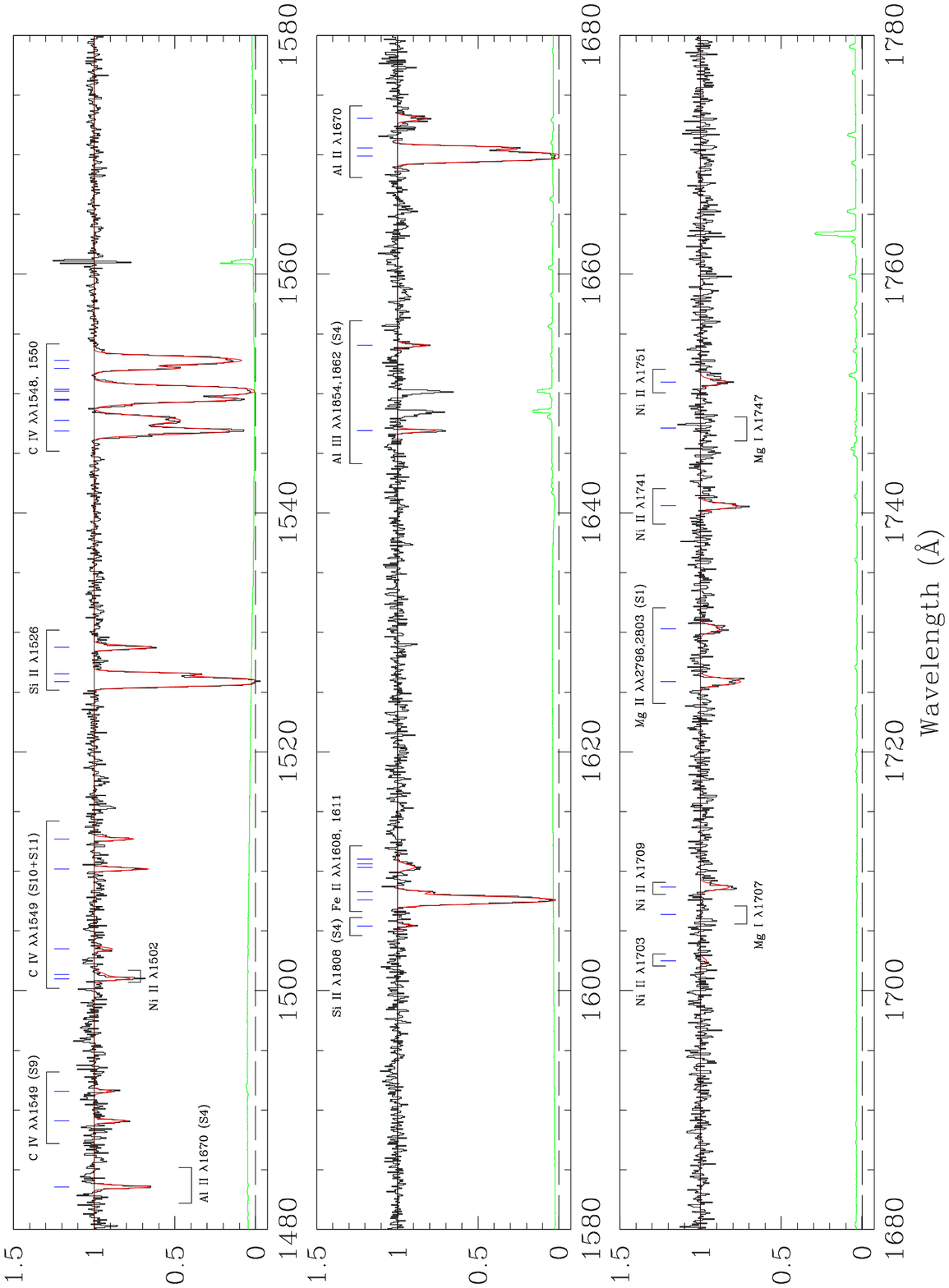}
\caption{continued \label{fig:plot_normalised_vpfit.2}}
\end{figure*}

\begin{figure*}
\figurenum{4d}
\includegraphics[width=0.95\textwidth]{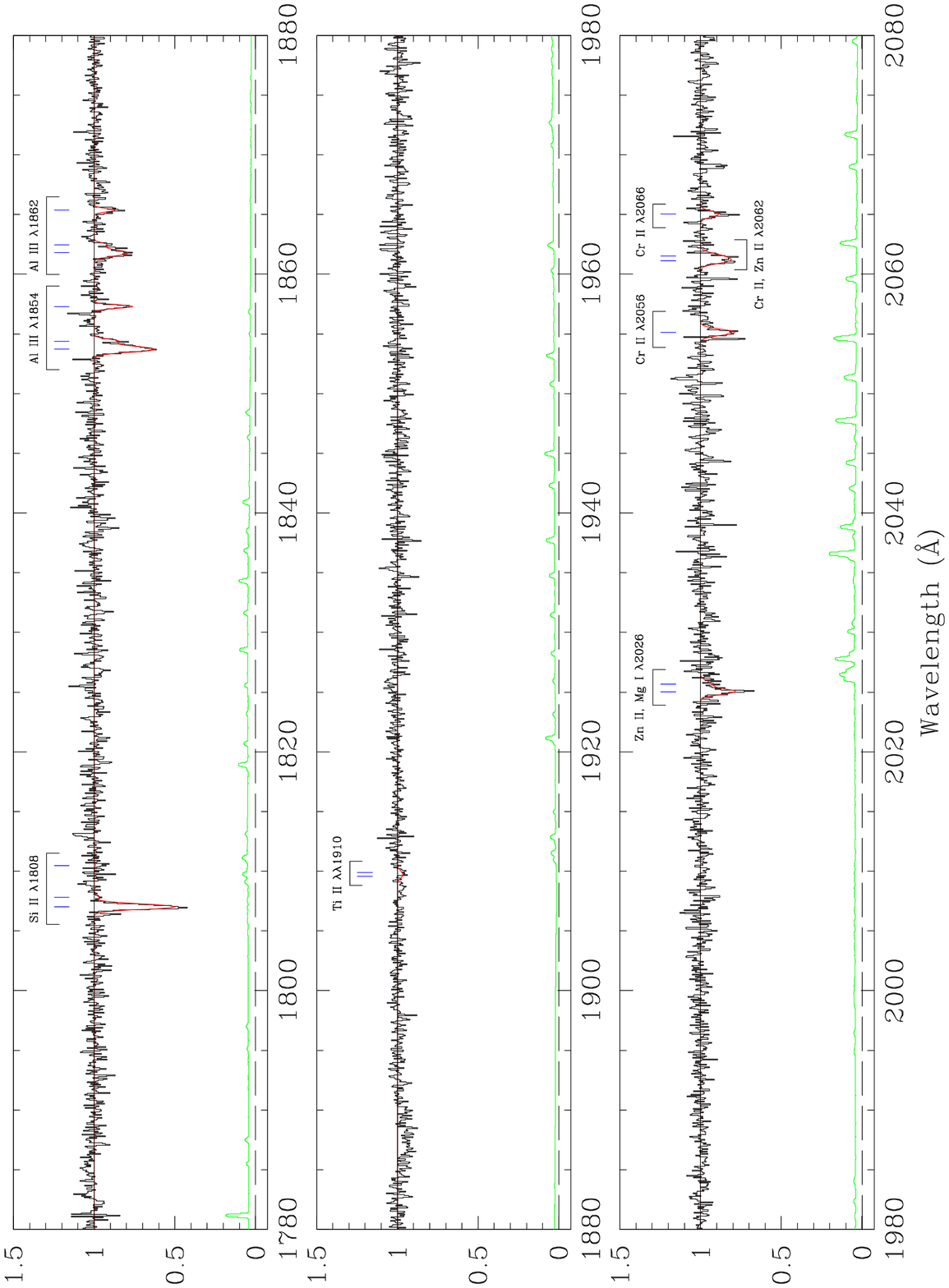}
\caption{continued \label{fig:plot_normalised_vpfit.3}}
\end{figure*}

\begin{figure*}
\figurenum{4e}
\includegraphics[width=0.95\textwidth]{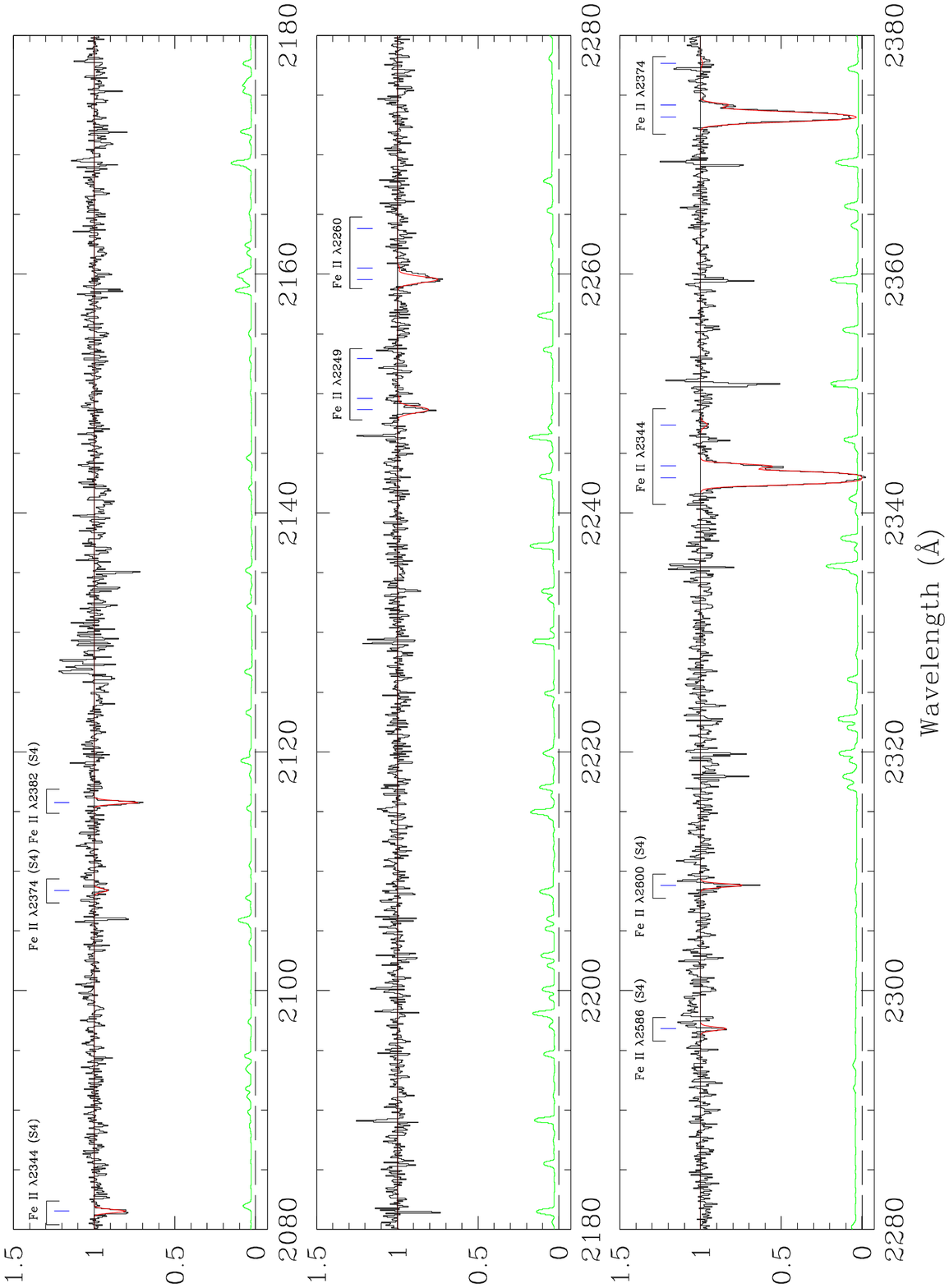}
\caption{continued \label{fig:plot_normalised_vpfit.4}}
\end{figure*}

\begin{figure*}
\figurenum{4f}
\includegraphics[width=0.95\textwidth]{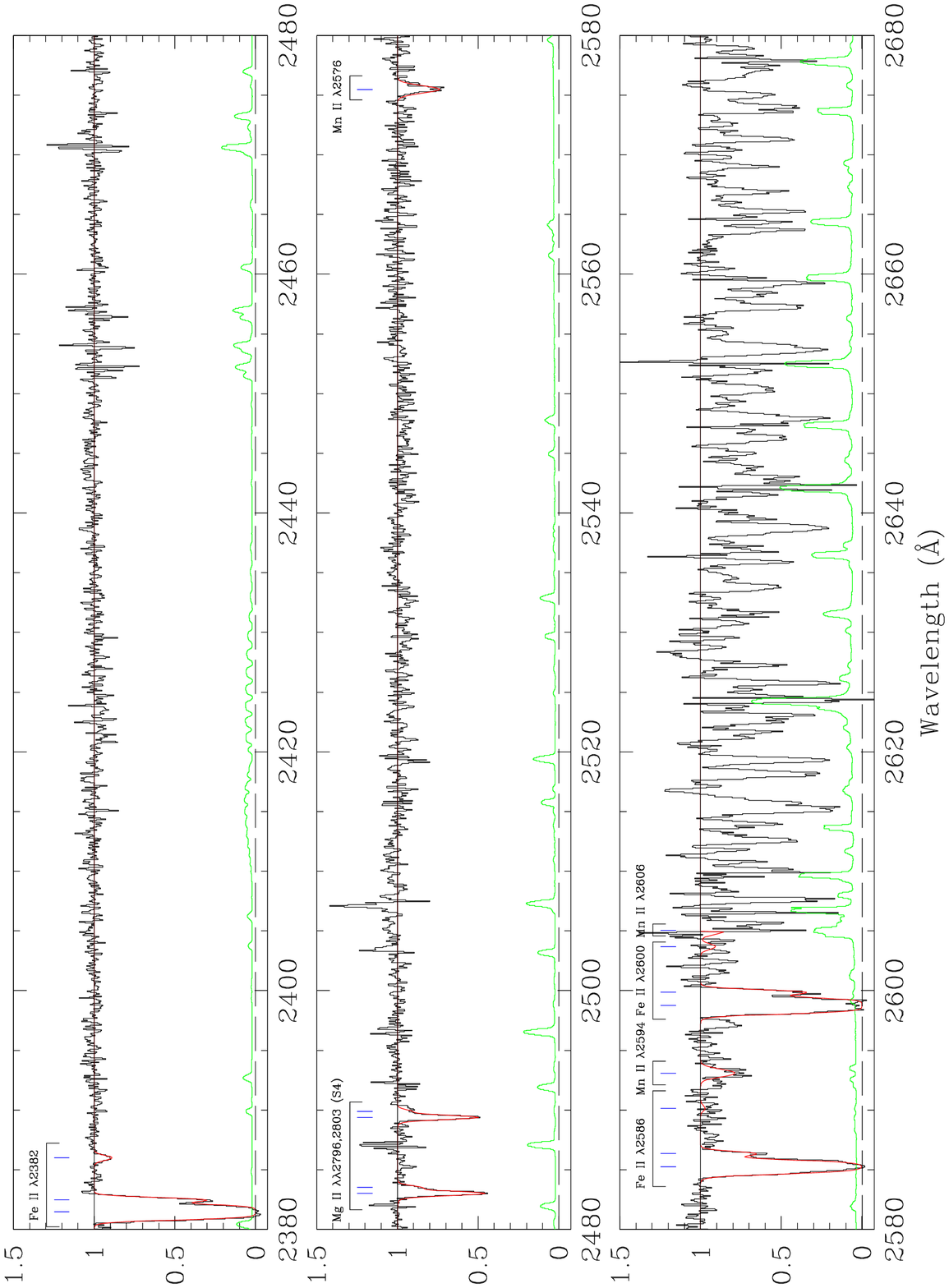}
\caption{continued \label{fig:plot_normalised_vpfit.5}}
\end{figure*}

\begin{figure*}
\figurenum{4g}
\includegraphics[width=0.95\textwidth]{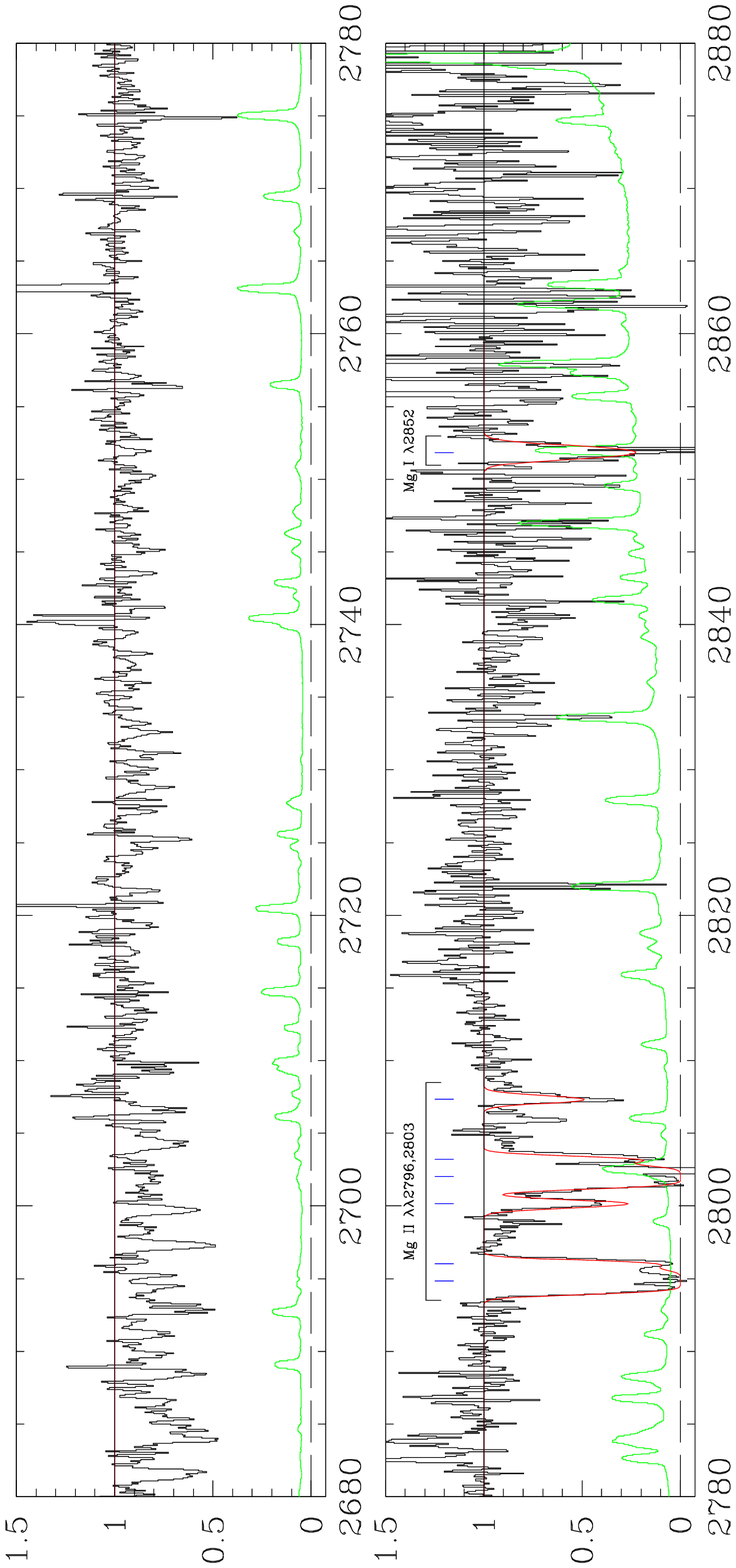}
\caption{continued \label{fig:plot_normalised_vpfit.6}}
\end{figure*}


\clearpage
\LongTables
\begin{deluxetable*}{lllcccccl}
\tablecolumns{9}
\tabletypesize{\small}
\tablecaption{\textsc{PDLA Absorption Lines}}
\tablehead{
  \multicolumn{1}{c}{Ion}
& \multicolumn{1}{c}{$\lambda_{\rm lab}$ \tablenotemark{a}}
& \multicolumn{1}{c}{$f$ \tablenotemark{a}}
& \multicolumn{1}{c}{$\Delta v$ \tablenotemark{b}}
& \multicolumn{1}{c}{$\lambda_{\rm obs}$\tablenotemark{c}}
& \multicolumn{1}{c}{$z_{\rm abs}$\tablenotemark{d}}
& \multicolumn{1}{c}{$W_0$\tablenotemark{e}}
& \multicolumn{1}{c}{$\sigma$\tablenotemark{f}}
& \multicolumn{1}{c}{Comments}\\
  \colhead{ }
& \multicolumn{1}{c}{(\AA)}
& \colhead{ }
& \multicolumn{1}{c}{(\kms)}
& \multicolumn{1}{c}{(\AA)}
& \colhead{ }
& \multicolumn{1}{c}{(\AA)}
& \multicolumn{1}{c}{(\AA)}
& \colhead{ }
}
\startdata
\\
\cutinhead {Low ionization species -- Blueshifted components}
\\
N\,{\sc i}    & 1134.656  & 0.0849  & $-$455 to $+$135 &  4053.10 & 2.5721 & 1.00 & 0.07 & Refers to N I $\lambda$1134 triplet. Upper limit\\ 
Fe\,{\sc ii}  & 1144.9379 & 0.0830  & $-$325 to  $+$50 &  4090.03 & 2.5723 & 0.54 & 0.04 & \\
P\,{\sc  ii}  & 1152.8180 & 0.245   & $-$250 to  $-$50 &  4117.37 & 2.5716 & 0.22 & 0.02 & Upper limit - blended with Ly$\alpha$\\
Si\,{\sc ii}  & 1190.4158 & 0.292   & $-$325 to  $+$50 &  4252.40 & 2.5722 & 1.05 & 0.02 & Upper limit as blended\\
Si\,{\sc ii}  & 1193.2897 & 0.582   & $-$325 to  $+$50 &  4262.58 & 2.5721 & 1.20 & 0.02 & Upper limit as blended\\ 
N\,{\sc  i}   & 1199.967  & 0.262   & $-$430 to $+$235 &  4286.71 & 2.5724 & 1.27 & 0.02 & Refers to N I $\lambda$1199 triplet \\
S\,{\sc ii}   & 1250.578  & 0.00543 & $-$250 to  $-$50 &  4467.09 & 2.5720 & 0.14 & 0.01 & \\ 
S\,{\sc ii}   & 1253.805  & 0.0109  & $-$250 to  $-$50 &  4478.53 & 2.5719 & 0.23 & 0.01 & \\ 
S\,{\sc ii}   & 1259.518  & 0.0166  & $-$250 to  $-$50 &  4499.20 & 2.5722 & 0.36 & 0.01 & Upper limit - blended with Si II $\lambda$ 1260\\ 
Si\,{\sc ii}  & 1260.4221 & 1.18    & $-$325 to  $+$50 &  4502.63 & 2.5723 & 1.15 & 0.02 & Upper limit - blended with S II $\lambda$ 1259\\ 
O\,{\sc i}    & 1302.1685 & 0.0480  & $-$325 to  $+$50 &  4651.62 & 2.5722 & 1.01 & 0.02 & \\ 
Si\,{\sc ii}  & 1304.3702 & 0.0863  & $-$325 to  $+$50 &  4659.47 & 2.5722 & 0.78 & 0.02 & Upper limit - blended with O I $\lambda$ 1302\\ 
Ni\,{\sc ii}  & 1317.217  & 0.0571  & $-$250 to  $-$50 &  4705.10 & 2.5720 & 0.09 & 0.01 & \\
C\,{\sc  ii}  & 1334.5323 & 0.128   & $-$325 to  $+$50 &  4767.44 & 2.5724 & 1.15 & 0.02 & \\ 
C\,{\sc  ii}* & 1335.7032 & 0.1278  & $-$250 to  $-$50 &  4771.00 & 2.5719 & 0.19 & 0.01 & \\ 
Ni\,{\sc ii}  & 1370.132  & 0.0588  & $-$250 to  $-$50 &  4894.39 & 2.5722 & 0.09 & 0.02 & \\ 
Si\,{\sc ii}  & 1526.7070 & 0.133   & $-$325 to  $+$50 &  5453.74 & 2.5722 & 1.06 & 0.01 & \\ 
Fe\,{\sc ii}  & 1608.4511 & 0.0577  & $-$325 to  $+$50 &  5745.66 & 2.5722 & 0.73 & 0.01 & \\ 
Al\,{\sc ii}  & 1670.7886 & 1.74    & $-$325 to  $+$50 &  5968.50 & 2.5723 & 1.17 & 0.02 & \\ 
Ni\,{\sc ii}  & 1709.6042 & 0.0324  & $-$250 to  $-$50 &  6106.92 & 2.5721 & 0.11 & 0.01 & \\ 
Ni\,{\sc ii}  & 1741.5531 & 0.0427  & $-$250 to  $-$50 &  6220.89 & 2.5720 & 0.14 & 0.01 & \\ 
Ni\,{\sc ii}  & 1751.9157 & 0.0277  & $-$250 to  $-$50 &  6258.17 & 2.5722 & 0.11 & 0.01 & \\ 
Si\,{\sc ii}  & 1808.0129 & 0.00208 & $-$250 to  $-$50 &  6458.18 & 2.5720 & 0.25 & 0.01 & \\ 
Ti\,{\sc ii}  & 1910.764  & 0.202   & $-$250 to  $-$50 & \nodata  &  \nodata & \phn$\leq 0.025$ \phle & \nodata & $5 \sigma$ upper limit \\
Zn\,{\sc ii}  & 2026.1370 & 0.501   & $-$250 to  $-$50 &  7237.65 & 2.5721 & 0.16 & 0.01 & Upper limit - blended with Mg I $\lambda$ 2026\\ 
Cr\,{\sc ii}  & 2056.2569 & 0.103   & $-$250 to  $-$50 &  7344.79 & 2.5719 & 0.14 & 0.03 & Noisy\\ 
Cr\,{\sc ii}  & 2062.2361 & 0.0759  & $-$250 to  $-$50 &  7366.69 & 2.5722 & 0.14 & 0.01 & Upper limit - blended with Zn II $\lambda$ 2062\\ 
Cr\,{\sc ii}  & 2066.1640 & 0.0512  & $-$250 to  $-$50 &  7379.98 & 2.5718 & 0.07 & 0.01 & \\ 
Fe\,{\sc ii}  & 2249.8768 & 0.00182 & $-$250 to  $-$50 &  8036.58 & 2.5720 & 0.10 & 0.01 & \\ 
Fe\,{\sc ii}  & 2260.7805 & 0.00244 & $-$250 to  $-$50 &  8075.80 & 2.5721 & 0.22 & 0.01 & \\ 
Fe\,{\sc ii}  & 2344.2139 & 0.114   & $-$325 to  $+$50 &  8373.97 & 2.5722 & 1.42 & 0.01 & \\ 
Fe\,{\sc ii}  & 2374.4612 & 0.0313  & $-$325 to  $+$50 &  8481.89 & 2.5721 & 0.98 & 0.01 & \\ 
Fe\,{\sc ii}  & 2382.7652 & 0.320   & $-$325 to  $+$50 &  8511.75 & 2.5722 & 1.70 & 0.02 & \\ 
Mn\,{\sc ii}  & 2576.877  & 0.361   & $-$250 to  $-$50 &  9204.81 & 2.5721 & 0.19 & 0.01 & \\ 
Fe\,{\sc ii}  & 2586.6500 & 0.0691  & $-$325 to  $+$50 &  9240.12 & 2.5722 & 1.40 & 0.02 & \\ 
Mn\,{\sc ii}  & 2594.499  & 0.280   & $-$250 to  $-$50 &  9267.56 & 2.5720 & 0.23 & 0.02 & Noisy\\ 
Fe\,{\sc ii}  & 2600.1729 & 0.239   & $-$325 to  $+$50 &  9288.40 & 2.5722 & 1.73 & 0.02 & Upper limit as blended\\ 
Mg\,{\sc ii}  & 2796.3543 & 0.6155  & $-$325 to  $+$50 &  9989.64 & 2.5724 & 2.27 & 0.03 & \\ 
Mg\,{\sc ii}  & 2803.5315 & 0.3058  & $-$325 to  $+$50 & 10015.06 & 2.5723 & 2.43 & 0.09 & Extremely noisy\\ 
Mg\,{\sc i}   & 2852.9631 & 1.83    & $-$325 to  $+$50 & 10191.84 & 2.5724 & 0.98 & 0.18 & Extremely noisy \\
\\
\cutinhead {Low ionization species -- Redshifted components}
\\
Si\,{\sc ii}  & 1193.2897 & 0.582   & $+$300 to $+$500 &  4270.39 & 2.5787 & 0.15 & 0.02 & \\    
N\,{\sc  i}   & 1199.967  & 0.262   & $+$195 to $+$685 &  4293.88 & 2.5783 & 0.25 & 0.02 & Refers to N I $\lambda$1199 triplet\\     
Si\,{\sc ii}  & 1260.4221 & 1.18    & $+$300 to $+$500 &  4510.79 & 2.5788 & 0.25 & 0.01 & \\     
C\,{\sc  ii}  & 1334.5323 & 0.128   & $+$300 to $+$500 &  4776.04 & 2.5788 & 0.28 & 0.01 & \\    
Si\,{\sc ii}  & 1526.7070 & 0.133   & $+$300 to $+$500 &  5463.67 & 2.5787 & 0.13 & 0.01 & \\     
Fe\,{\sc ii}  & 1608.4511 & 0.0577  & $+$300 to $+$500 &  5755.34 & 2.5782 & 0.07 & 0.01 & Upper limit - blended with Fe II $\lambda$1611\\
Al\,{\sc ii}  & 1670.7886 & 1.74    & $+$300 to $+$500 &  5979.17 & 2.5787 & 0.10 & 0.01 & \\     
Fe\,{\sc ii}  & 2382.7652 & 0.320   & $+$300 to $+$500 &  8527.38 & 2.5788 & 0.06 & 0.01 & \\     
Mg\,{\sc ii}  & 2796.3543 & 0.6155  & $+$300 to $+$500 & 10008.06 & 2.5790 & 0.54 & 0.03 & Upper limit\\     
Mg\,{\sc ii}  & 2803.5315 & 0.3058  & $+$300 to $+$500 & 10033.37 & 2.5788 & 0.46 & 0.03 & Upper limit\\     
\\
\cutinhead {High ionization species -- Blueshifted components}
\\
Si\,{\sc iii} & 1206.500  & 1.63    & $-$325 to  $+$50 &  4310.11 & 2.5724 & 1.23 & 0.02 & \\ 
N\,{\sc v}    & 1238.821  & 0.1560  & $-$400 to $-$100 &  4423.54 & 2.5708 & 0.42 & 0.01 & \\  
N\,{\sc v}    & 1242.804  & 0.07770 & $-$400 to $-$100 &  4437.84 & 2.5708 & 0.22 & 0.01 & Upper limit - blended with N V $\lambda$1238\\ 
Si\,{\sc iv}  & 1393.7602 & 0.513   & $-$325 to  $+$50 &  4979.13 & 2.5724 & 0.57 & 0.02 & \\ 
Si\,{\sc iv}  & 1402.7729 & 0.254   & $-$325 to  $+$50 &  5011.57 & 2.5726 & 0.29 & 0.02 & \\ 
C\,{\sc iv}   & 1548.204  & 0.1899  & $-$400 to  $+$50 &  5529.73 & 2.5717 & 0.99 & 0.00 & \\ 
C\,{\sc iv}   & 1550.781  & 0.09475 & $-$400 to  $+$50 &  5539.46 & 2.5720 & 1.36 & 0.00 & Upper limit - blended with C IV $\lambda$1548\\
Al\,{\sc iii} & 1854.7184 & 0.559   & $-$325 to  $+$50 &  6625.72 & 2.5724 & 0.28 & 0.01 & \\
Al\,{\sc iii} & 1862.7910 & 0.278   & $-$325 to  $+$50 &  6653.82 & 2.5720 & 0.19 & 0.01 & \\ 
\\
\cutinhead {High ionization species -- Redshifted components}
\\
Si\,{\sc iii} & 1206.500  & 1.63    & $+$300 to $+$500 &  4317.79 & 2.5788 & 0.42 & 0.02 & Upper limit\\ 
N\,{\sc v}    & 1238.821  & 0.1560  & $+$150 to $+$600 &  4432.62 & 2.5781 & 0.42 & 0.01 & \\   
N\,{\sc v}    & 1242.804  & 0.07770 & $+$150 to $+$600 &  4447.17 & 2.5783 & 0.28 & 0.01 & \\   
Si\,{\sc iv}  & 1393.7602 & 0.513   & $+$150 to $+$600 &  4987.55 & 2.5785 & 0.59 & 0.02 & \\   
Si\,{\sc iv}  & 1402.7729 & 0.254   & $+$150 to $+$600 &  5019.95 & 2.5786 & 0.32 & 0.02 & \\  
C\,{\sc iv}   & 1548.204  & 0.1899  & $+$150 to $+$600 &  5539.54 & 2.5780 & 1.36 & 0.00 & Upper limit - blended with C IV $\lambda$1550\\   
C\,{\sc iv}   & 1550.781  & 0.09475 & $+$150 to $+$600 &  5549.17 & 2.5783 & 0.87 & 0.00 & \\   
Al\,{\sc iii} & 1854.7184 & 0.559   & $+$300 to $+$500 &  6638.38 & 2.5792 & 0.06 & 0.01 & \\  
Al\,{\sc iii} & 1862.7910 & 0.278   & $+$300 to $+$500 &  6666.32 & 2.5787 & 0.08 & 0.01 & \\   
\enddata 
\tablenotetext{a}{Vacuum wavelengths and $f$-values from 
the compilation by \cite{morton03} with recent
updates by \cite{jenkins06}.}
\tablenotetext{b}{Velocity range, relative to $ z_{\rm sys}  = 2.57393$, 
over which the equivalent widths were measured.}
\tablenotetext{c}{Centroid wavelength of the absorption line in the observed frame.}
\tablenotetext{d}{Measured absorption redshift.}
\tablenotetext{e}{Rest-frame equivalent width.}
\tablenotetext{f}{$1\sigma$ random error for the rest-frame equivalent width.}
\tablecomments{The systematic errors are comparable to the random errors.}
\label{tab:table_dla_ew}
\end{deluxetable*}
\bigskip

\section{The Kinematics of the PDLA}\label{sec:kinematics}

We examine the kinematics of the DLA
in Figure~\ref{fig:plot_velocity_structure} by comparing the
absorption line profiles of ions in different ionization
stages; the zero point of the velocity scale is at 
$z_{\rm sys} = 2.57393$.
The first thing to note is that evidently there are two 
main clumps of absorption (in velocity space),
one receding from the quasar while the other is 
apparently approaching it.
The blueshifted absorption ranges in velocity
from $+50$ to $-325$\kms, while the redshifted
components span from $+150$ to $+600$\kms.
Such velocities are of the same order of magnitude as those 
commonly seen in galactic-scale outflows in star-forming
galaxies at redshifts $z \simeq 2 - 3$
\citep[see \eg][]{pettini02, erb04},
raising the possibility that we are witnessing
motions of gas associated with the host galaxy of
the quasar.
Second, while the low and high ions 
span broadly similar velocity ranges, 
there are differences in their detailed 
kinematics, both in the relative strengths of
different components and in the velocity
where the optical depth is largest. We now 
discuss these variations in more detail.

In Figure~\ref{fig:plot_velocity_structure} 
we have chosen 
Si\,{\sc ii}~$\lambda 1526$ 
(at the top in both columns)
as an example
of the absorption by gas that is predominantly
neutral (Si\,{\sc ii} being the dominant ion stage
of silicon in H\,{\sc i} regions). Moving down
each column, we show gas that is increasingly more highly
ionized from Al{\scriptsize ~III} and Si{\scriptsize ~III}, 
to C{\scriptsize ~IV} and Si{\scriptsize ~IV},
and finally N{\scriptsize ~V}.  The first four panels
in the right-hand column are particularly useful as
they illustrate the evolution in velocity structure 
with ionization state for a {\em
single\/} element, silicon.

Transitions from low ions such as  Si\,{\sc ii}~$\lambda 1526$ 
show a main component (C1), with peak optical depth at 
$\sim -$162\kms, blended with weaker one (C2) at $\sim -$35\kms; 
a third, weakest, component (C3) is centred at $\sim
+$406\kms. While the Si\,{\sc iii}~$\lambda 1206$ line is too
saturated to discern much information on its velocity structure, 
the weak Al\,{\sc iii}~$\lambda \lambda 1854,1862$ doublet 
shows that the strengths of components C2 and C3 
relative to C1 increase in  
moderately ionized gas, although there is
little difference ($<20$\kms) in their relative velocities 
compared to Si\,{\sc ii}~$\lambda 1526$.
The trend continues as we move to
Si\,{\sc iv}, with C2 and C3 increasing further in strength
compared to C1. Furthermore, a fourth component (C4) 
appears at $\sim+$257\kms. 
The C\,{\sc iv}~$\lambda\lambda 1548,1550$ lines 
show the same pattern, even though the individual components
are more difficult to disentangle given that the doublet
lines' separation, $\Delta v = 499$\kms, is comparable to
the velocity differences between components.
Finally, the N\,{\sc v}~$\lambda \lambda
1238,1242$ doublet reveals even more pronounced differences in the
properties of the very highly ionized gas, with
a {\em single} blueshifted absorber (rather than two components)
at $\sim -$250\kms, and two redshifted components
of differing widths but centred at similar velocities
near $\sim+$357\kms, intermediate between those
of C3 and C4.

In summary, it appears that the bulk of the neutral gas resides in
components C1 and C2, which are receding from the quasar, whilst
ionized gas is more significant in the components that are approaching
it (C3 and C4). In Table~\ref{tab:table_dla_ew} we have listed
separately the equivalent widths of absorption lines in the
blueshifted and redshifted components; the velocity ranges over
which the equivalent width integration was carried out are given in
column (4) and are indicated by vertical dashed lines in
Figure~\ref{fig:plot_velocity_structure}.

\addtocounter{figure}{+1}
\begin{figure*}
\begin{center}
\includegraphics[width=0.8\textwidth]{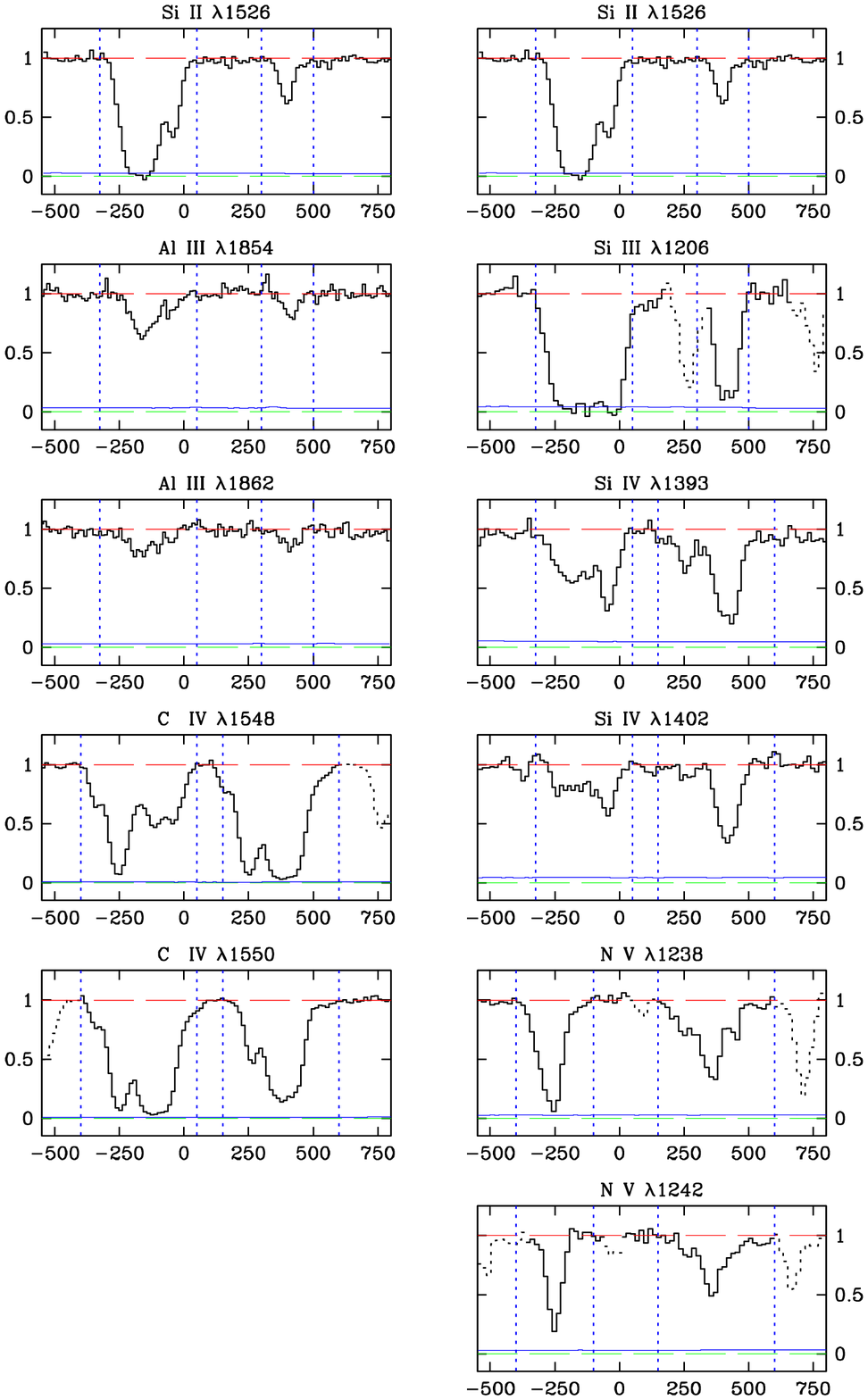}
\caption{The velocity structure of the PDLA's absorbing gas and its
dependence on ionization level. Each plot shows the normalized
absorption profile for a particular species ({\em histogram}); 
the line near the zero level is the corresponding
$1 \sigma$ error spectrum. The $x$-axis is
velocity (in \kms) relative to $ z_{\rm sys} = 2.57393$;
the $y$-axis is residual flux. 
Blending with lines from other species is flagged
by a dotted line type. In addition, note that the redshifted
component of \civa\ is blended with the blueshifted component of \civb,
since the doublet lines' separation is 499\,\kms.
The dashed vertical lines indicate the two wavelength ranges 
over which the equivalent widths listed in 
Table~\ref{tab:table_dla_ew} were measured.}
\label{fig:plot_velocity_structure}
\end{center}
\end{figure*}

\begin{deluxetable}{ccccc}
\tablewidth{0pt}
\tablecaption{\textsc{Absorption Components of Low Ion
Transitions}\label{tab:abs_comp}}
\tablehead{
  \colhead{Component}
& \colhead{$z_{\rm abs}$}
& \colhead{$\Delta v$\tablenotemark{a}}
& \colhead{$b$}
& \colhead{Fraction\tablenotemark{b}}\\
   \colhead{}
& \colhead{}
& \colhead{(\kms)}
& \colhead{(\kms)}
& \colhead{}
}
\startdata
\\
C1 & 2.57200 &   $-162$   & 41.6  &  0.96 \\
C2 & 2.57351 &   $-35$    & 14.4  &  0.03 \\
C3 & 2.57877 &   $+406$   & 26.6  &  0.01 \\
\enddata
\tablenotetext{a}{Velocity difference from $z_{\rm sys} = 2.57393$.}
\tablenotetext{b}{Fraction of the total column density of Si\,{\sc ii}.}
\end{deluxetable}
\bigskip

\section{The Chemical Properties of the PDLA}\label{sec:abundances}

\subsection{Ion Column Densities}\label{subsec:ion_col_densities}

We derived ion column densities by fitting the absorption lines with
multiple Voigt profiles generated by VPFIT (version 8.02), a software
package that uses $\chi^2$ minimisation to deduce the values of
redshift $z$, column density $N$ (cm$^{-2}$), and Doppler parameter
$b$ (\kms) that best reproduce the observed absorption line
profiles.  VPFIT\footnote{VPFIT is available from
http://www.ast.cam.ac.uk/\textasciitilde rfc/vpfit.html.} takes into
account the instrumental broadening function (which in the present
case is well approximated by a Gaussian with FWHM = 45.5\kms---see
\S\ref{sec:observations}) in its $\chi^2$ minimisation and error
evaluation, and is specifically designed to fit blends of lines
(provided there are sufficient constraints). We used the compilation of
laboratory wavelengths and $f$-values by \cite{morton03} with recent
updates by \cite{jenkins06}.  The best fitting line profiles returned
by VPFIT are compared with the observed ones in
Figures~\ref{fig:plot_normalised_vpfit.0}--\ref{fig:plot_normalised_vpfit.6}.

\begin{deluxetable}{lcccc}
\tablewidth{0pt}
\tablecaption{\textsc{Column Densities for Low Ionization Species}\label{tab:abs_comp_coldensity}}
\tablehead{
  \colhead{Ion}
& \colhead{$\log N$(C1)}
& \colhead{$\log N$(C2)}
& \colhead{$\log N$(C3)}
& \colhead{$\log$ [$N$(C1) + $N$(C2)]}\\
  \colhead{}
& \colhead{(cm$^{-2}$)}
& \colhead{(cm$^{-2}$)}
& \colhead{(cm$^{-2}$)}
& \colhead{(cm$^{-2}$)}
}
\startdata
C\,{\sc ii}  & $>$16.336\phl & $>$14.836\phl & 14.421  & $>$16.350\phl \\
N\,{\sc i}   &    14.987     &    13.463     & 13.810  &    15.000 \\
O\,{\sc i}   & $>$16.890\phl & $>$14.659\phl & \nodata & $>$16.893\phl \\ 
Mg\,{\sc ii} &    15.486     &    13.986     & 13.361  &    15.500 \\
Al\,{\sc ii} & $>$14.546\phl & $>$13.315\phl & 12.268  & $>$14.571\phl \\
Si\,{\sc ii} &    15.783     &    14.234     & 13.778  &    15.795 \\
P\,{\sc ii}  & $<$13.686\phl & $<$12.200\phl & \nodata & $<$13.700\tablenotemark{a}\phd \\
S\,{\sc ii}  &    15.380     &    \nodata    & \nodata &    15.380 \\ 
Ti\,{\sc ii} & $<$12.585\phl &    \nodata    & \nodata & $<$12.585\phl \\
Cr\,{\sc ii} &    13.577     &    \nodata    & \nodata &    13.577 \\
Mn\,{\sc ii} &    13.041     &    \nodata    & \nodata &    13.041 \\
Fe\,{\sc ii} &    15.229     &    13.744     & 12.599  &    15.243 \\
Ni\,{\sc ii} &    14.078     &    \nodata    & \nodata &    14.078 \\
Zn\,{\sc ii} &    12.898     &    \nodata    & \nodata &    12.898 \\
\enddata

\tablenotetext{a}{As P{\scriptsize\,II}~$\lambda1152$ was blended with
a \lya\ line, we derived upper limits to the column densities of C1
and C2 under the assumption that their ratio is similar to that for
Fe{\scriptsize\,II}.\\}
\tablecomments{The table presents the ion column densities measured
for the three absorption components listed in Table~\ref{tab:abs_comp}.  We
attempted to fit all three components for each ion; in practice, for
species which have only weak lines (\eg\ S{\scriptsize\,II} and
Zn{\scriptsize\,II}), C2 could not be distinguished from C1, so that a
single component was fitted.  In these cases, the weaker third
component, C3, was not detected.  We were also unable to fit C3 in
O{\scriptsize\,I}, due to the blending of the
O{\scriptsize\,I}~$\lambda 1302$ line with the main component of
Si{\scriptsize\,II}~$\lambda 1304$. For reasons explained in the text,
this third component was not used in our element abundance
analysis. Instead we used the summed column density of the other two
components, as given in the final column of the table.}
\end{deluxetable}
\bigskip

In agreement with our description of the line profiles
in \S\ref{sec:kinematics}, we
found that a three component model was necessary, and sufficient, to
characterise the absorption from the low ionization gas. 
The kinematic properties of the species that are dominant
in H\,{\sc i} regions are remarkably uniform:  on a first
pass in the fitting procedure---when each species is fitted
independently of all the others---we found
that the redshifts and $b$ values returned  
for each component by VPFIT were in good agreement 
from ion to ion.
We therefore subsequently tied these parameters together
for all the low ionization species, 
so that each of the three components was characterized by a
single value of redshift $z$ and Doppler parameter $b$,
listed in Table~\ref{tab:abs_comp}.
VPFIT was able to converge to an excellent fit for the first two
absorption components (C1 \& C2). 
The fitting of C3 (the redshifted component)
was however much poorer, 
with the absorption being overpredicted for the stronger
lines. This is a classic signature of partial coverage, 
which occurs when the foreground gas does not cover 
all of the continuum source,
suggesting that C3 is physically close to the quasar.
We return to this point in \S\ref{sec:partial_coverage} below.

Table~\ref{tab:abs_comp_coldensity} lists the column densities fitted,
for each component, for species that are the dominant ion stages of
their respective elements in H\,{\sc i} regions. The total column
densities, derived by adding together components C1 and C2, are
summarised in the final column of the table and again in column~2 of
Table~\ref{tab:abundances}. Inclusion of component C3, which in any
case is likely to consist mostly of ionized gas, would increase the
column densities of species such as Si\,{\sc ii} and Fe\,{\sc ii} by
less than 1\% (see discussion in \S\ref{sec:partial_coverage}).  The
values for C\,{\sc ii}, O\,{\sc i} and Al\,{\sc ii} are lower limits
because the corresponding absorption lines are saturated (see
Figures~\ref{fig:plot_normalised_vpfit.1} and
\ref{fig:plot_normalised_vpfit.2}); the limits quoted were derived
under the assumption that the model parameters listed in
Table~\ref{tab:abs_comp} apply to the saturated lines too.  We
consider the column density of P\,{\sc ii} to be an upper limit
because of blending with a \lya\ forest line
(Figure~\ref{fig:plot_normalised_vpfit.0}), while Ti\,{\sc ii} may be
marginally detected at the $5 \sigma$ level in the $\lambda\lambda
1910.6, 1910.9$ doublet (which is unresolved in our data).  The value
for Mg\,{\sc ii} is uncertain (as is usually the case in DLAs),
because the $\lambda\lambda 2796,2803$ doublet lines are strongly
saturated, while those of the $\lambda\lambda 1239.9,1240.4$ doublet
are (a) blended with N\,{\sc v}~$\lambda\lambda 1238,1242$, and (b) so
weak as to be very sensitive to the exact placement of the continuum
level in the red wing of the \lya+N\,{\sc v} broad emission line.

\subsection{Element Abundances}\label{subsec:abundances}

\begin{figure}
\centerline{\includegraphics[width=0.95\columnwidth]{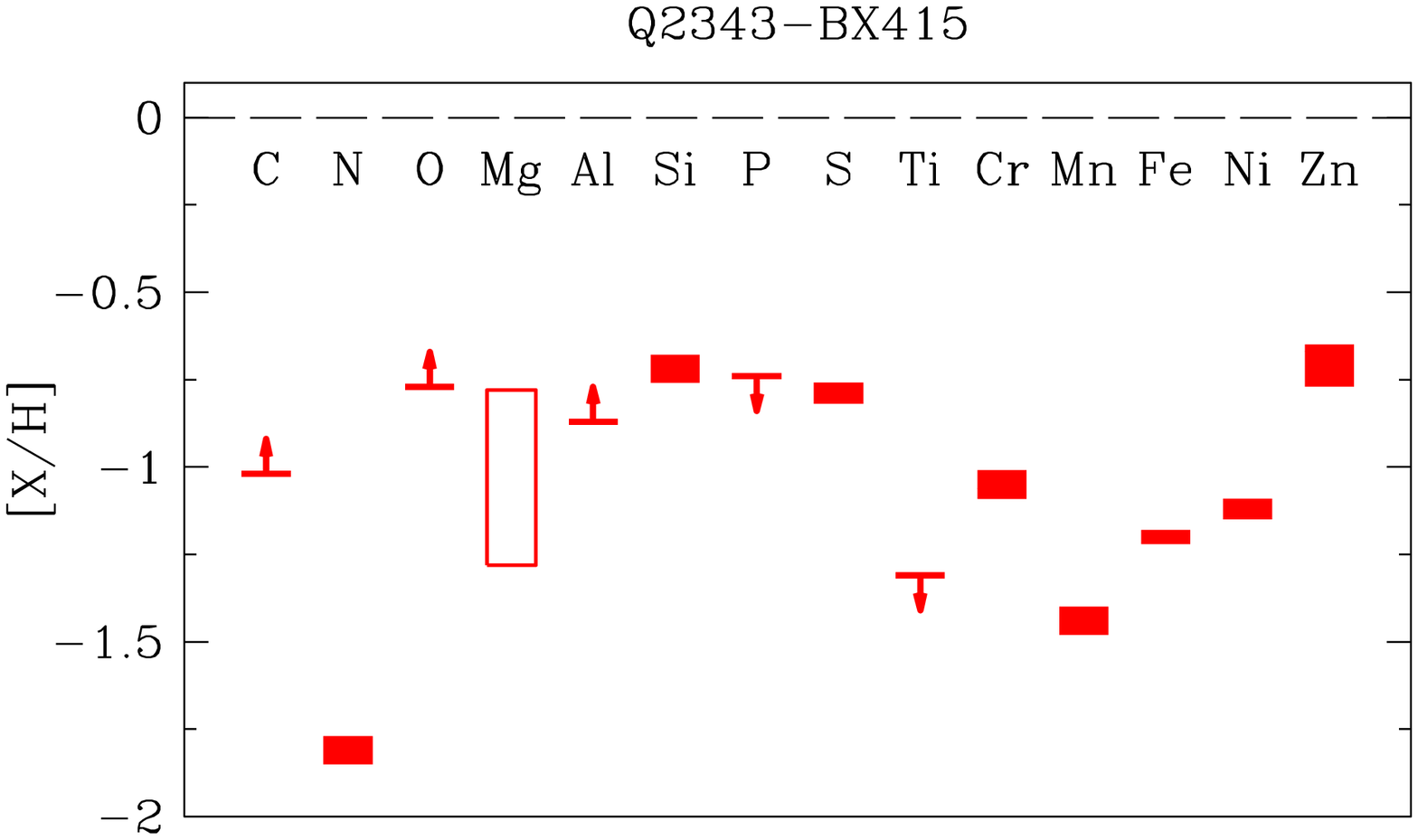}}
\caption{Element abundances of the PDLA in Q2343--BX415, relative to the
solar abundance scale of \cite{lodders03}. Abundance determinations
are shown by boxes, where the height of the box reflects the error
estimate from VPFIT, while arrows indicate upper and lower
limits. For reasons explained in the text, the measurement of
the Mg abundance is uncertain, 
and is indicated here by an empty box.\label{fig:plot_abundances}}
\end{figure}

Abundances (or limits) 
relative to the solar scale \citep{lodders03}
are given in the last column of Table~\ref{tab:abundances}
for 14  elements in the PDLA under the usual
assumption that
\begin{equation}\label{eq:ionization_assumption}
\log ({\rm X / H}) = \log ({\rm X}^i / {\rm H}^0),
\label{eq:ion_approx}
\end{equation}
where X$^i$ is the dominant state of element X in neutral gas. 
While this approximation has been shown to be appropriate 
for most intervening DLAs \citep{vladilo01}, this may not
be the case here, given the proximity of the PDLA to
Q2343--BX415. We address this issue in detail in
\S\ref{sec:ionization_corrections} and conclude that
the approximation in equation~(\ref{eq:ion_approx}) is in fact
valid in the present case too. 
An additional consideration concerning the accuracy of the
abundance measurements in Table~\ref{tab:abundances}
is that we have neglected the contribution of molecular
hydrogen to the total value of $N$(H). We have no way 
of estimating this correction directly, as our ESI spectrum of 
Q2343--BX415 does not extend below 1130\,\AA\ in the rest
frame, and therefore does not cover any H$_2$ absorption lines; 
however, unless this DLA is very unusual, the fraction 
of molecular gas is unlikely to be higher than  2\% \citep{petitjean06}.

\begin{deluxetable}{lcccc}
\tabletypesize{\small}
\tablewidth{0pt}
\tablecaption{\textsc{Element Abundances}\label{tab:abundances}}
\tablehead{
\multicolumn{1}{l}{Ion} 
& \multicolumn{1}{c}{$\log N$ (cm$^{-2}$)}
& \multicolumn{1}{c}{$\log$ (X/H)}
& \multicolumn{1}{c}{$\log$ (X/H)$_{\odot}$\tablenotemark{a}}
& \multicolumn{1}{c}{[X/H]$_{\rm PDLA}$\tablenotemark{b}}
}
\startdata
H\,{\sc i}   & \phm{$>$}20.98 & \phm{$>$}\nodata & \nodata & \nodata\\
C\,{\sc ii}  &       $>$16.35 &       $>$$-$4.63 & $-$3.61 & $>-1.02$\\
N\,{\sc i}   & \phm{$>$}15.00 & \phm{$>$}$-$5.98 & $-$4.17 & $-1.81\pm0.04$\phm{:}\\
O\,{\sc i}   &       $>$16.89 &       $>$$-$4.08 & $-$3.31 & $>-0.77$\\
Mg\,{\sc ii} & \phm{$>$}15.50 & \phm{$>$}$-$5.48 & $-$4.45 & $-1.03\pm0.25$:\\
Al\,{\sc ii} &       $>$14.57 &       $>$$-$6.41 & $-$5.54 & $>-0.87$\\
Si\,{\sc ii} & \phm{$>$}15.79 & \phm{$>$}$-$5.18 & $-$4.46 & $-0.72\pm0.04$\phm{:}\\
P\,{\sc ii}  &       $\leq 13.70$ &       $\leq  -7.28$ & $-6.54$ & $\leq -0.74$\\
S\,{\sc ii}  & \phm{$>$}15.38 & \phm{$>$}$-$5.60 & $-$4.81 & $-0.79\pm0.03$\phm{:}\\
Ti\,{\sc ii} &      $\leq 12.59$ & $\leq  -8.39$ & $-7.08$ & $\leq -1.31$\\
Cr\,{\sc ii} & \phm{$>$}13.58 & \phm{$>$}$-$7.40 & $-$6.35 & $-1.05\pm0.04$\phm{:}\\
Mn\,{\sc ii} & \phm{$>$}13.04 & \phm{$>$}$-$7.94 & $-$6.50 & $-1.44\pm0.04$\phm{:}\\
Fe\,{\sc ii} & \phm{$>$}15.24 & \phm{$>$}$-$5.73 & $-$4.53 & $-1.20\pm0.02$\phm{:}\\
Ni\,{\sc ii} & \phm{$>$}14.08 & \phm{$>$}$-$6.90 & $-$5.78 & $-1.12\pm0.03$\phm{:}\\
Zn\,{\sc ii} & \phm{$>$}12.90 & \phm{$>$}$-$8.08 & $-$7.37 & $-0.71\pm0.06$\phm{:}\\
\enddata
\tablenotetext{a}{Solar abundance scale from \cite{lodders03}.}
\tablenotetext{b}{[X/H]$_{\rm PDLA} = \log {\rm (X/H)} - \log {\rm (X/H)}_{\odot}$. 
The errors quoted are from the VPFIT error estimates associated with 
each value of $\log N{\rm (X)}$ and 
do not include the uncertainty in the determination of 
$\log N{\rm (H)}$. The latter would result in the same offset for
all elements.}
\tablecomments{The H\,{\sc i} column density was measured by fitting the
damping wings of the \lya\ absorption line (see \S\ref{subsec:p-dla}). 
Column densities for all the other ions were determined from Voigt profile
fitting with VPFIT (see Table~\ref{tab:abs_comp_coldensity}).}
\end{deluxetable}
\bigskip

The element abundance pattern of the PDLA in Q2343--BX415
is illustrated in Figure~\ref{fig:plot_abundances}. 
The interpretation of such abundance measurements has been discussed
extensively in the literature \citep[see, for example,][]{prochaska03, wolfe05};
suffice here to say that the observed pattern of element abundances
reflects both the previous history of stellar nucleosynthesis and 
the depletion of refractory elements onto interstellar dust grains.
Combining our knowledge of depletions in the interstellar
gas of the Milky Way and of element abundances in different
stellar populations, it is usually possible to disentangle the two effects.
From consideration of the results in 
Figure~\ref{fig:plot_abundances} we draw the following conclusions:

(1) The overall metallicity of the PDLA is 
$Z_{\rm PDLA} \simeq 1/5 -1/6 \,Z_{\odot}$,
since for the elements that are not usually depleted onto dust we have
${\rm [O, S, Zn/H]} \simeq -0.7$ to $-0.8$.

(2)  There appears to be no intrinsic enhancement of alpha-capture
elements relative to the Fe-group, if we take the abundance of the
undepleted S and Zn to be representative of, respectively, 
these two groups of elements:
${\rm [S/Zn]} \simeq -0.1$.

(3) The underabundances of Ti, Cr, Fe and Ni relative to Zn (all
Fe-peak elements) are most naturally explained as the result
of moderate dust depletions, analogous to those observed
in diffuse clouds in the halo of the Milky Way
\citep{savage96}. At these modest depletion levels
(${\rm [Fe/Zn]} \simeq  -0.5$), the finding that Si is not 
depleted while Mg apparently is
(${\rm [Si/S]} \simeq  +0.1$ while 
${\rm [Mg/S]} \simeq  -0.24$---all three are 
alpha-capture elements) is also consistent with 
analogous measurements in the Galactic ISM
\citep[see Table 6 of][]{savage96}.

(4) The enhanced underabundance of Mn probably results from a combination
of dust depletion and a genuine decrease in the intrinsic [Mn/Fe] ratio
at lower metallicities \citep{pettiniellison00}.

(5) The element with the lowest abundance is N.
While somewhat extreme, the low (N/O) 
ratio ($\log {\rm (N/O)} \lesssim -1.9$ or
${\rm [N/O]} \lesssim -1.0$) is nevertheless
in the sense expected from the steep dependence
of the N production on stellar metallicity
\citep[see, for example, the discussion by][]{pettiniellison02}.

\begin{figure}
\medskip
\centerline{\includegraphics[angle=270,width=0.9\columnwidth]{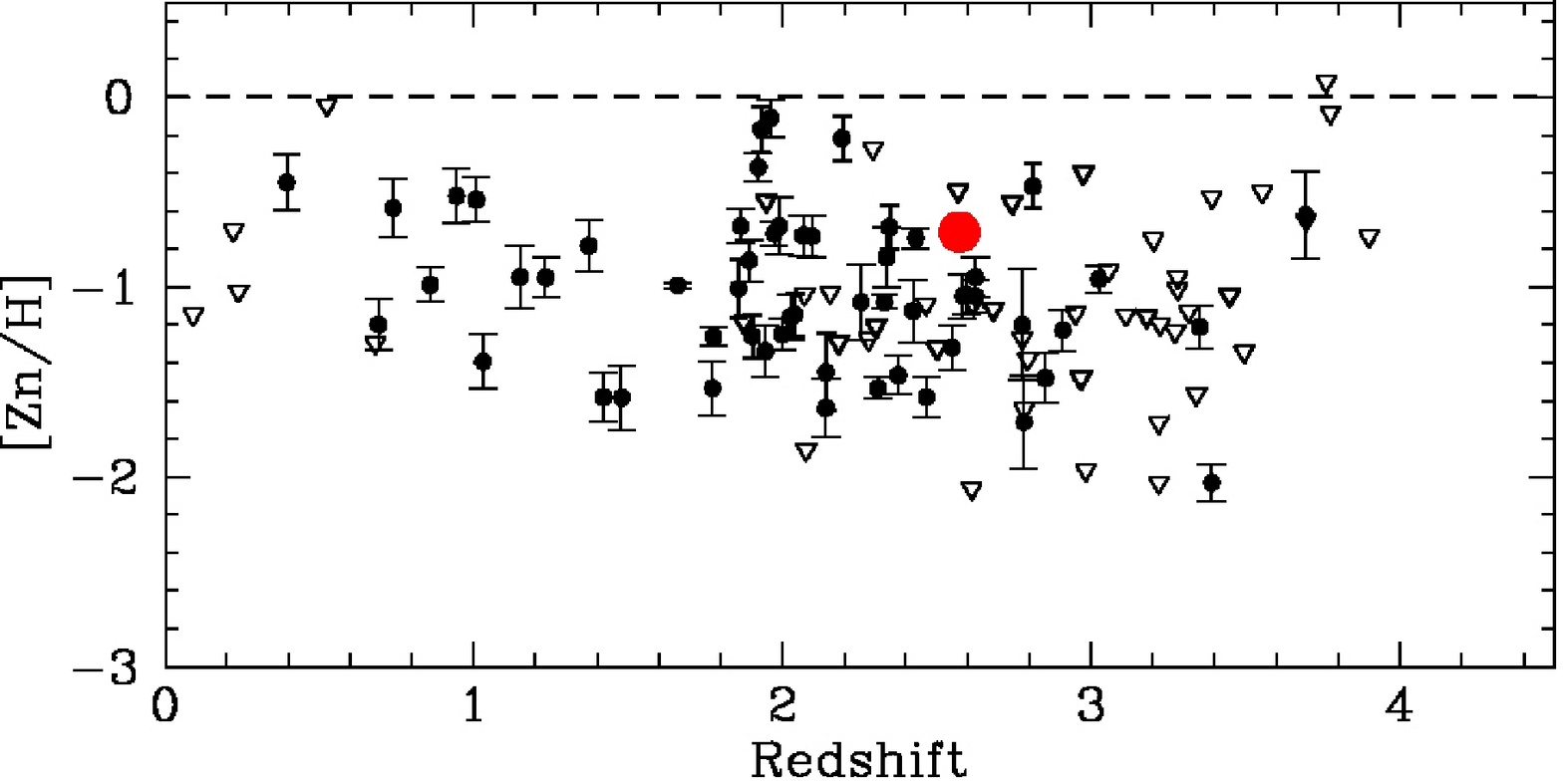}}
\smallskip
\caption{Metallicity of DLAs measured via the [Zn/H] ratio. The PDLA studied
here is shown by the large circle; the other points are from the compilation
of over 100 DLAs by \cite{akerman05}. Filled circles are cases where the generally
weak Zn\,{\sc ii} absorption lines have been detected, while 
open triangles are $3 \sigma$ upper limits corresponding to non-detections.\\\label{fig:plot_Zn_abund}}
\end{figure}

\begin{figure}
\medskip
\centerline{\includegraphics[angle=270,width=0.9\columnwidth]{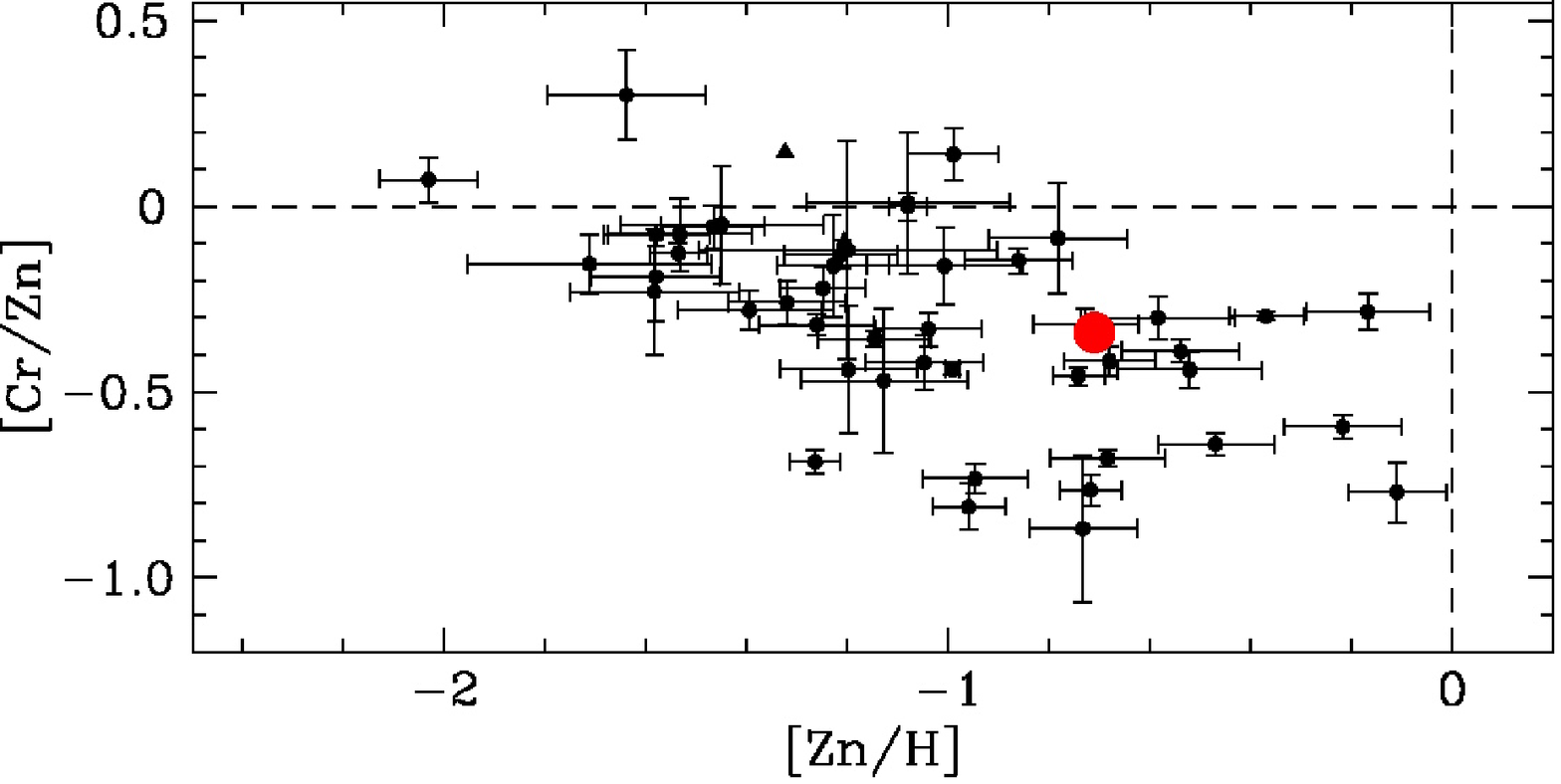}}
\smallskip
\caption{The dependence of the [Cr/Zn] ratio on metallicity in DLAs.
The data are from \cite{akerman05} except for the large circle which refers
to the PDLA in Q2343--BX415. The [Cr/Zn] ratio gives an indication
of the extent to which refractory elements are incorporated into dust
grains.\\\label{fig:plot_ZnCr}}
\end{figure}

However, the most significant conclusion for the 
purposes of the present study is that findings (1)--(5)
are entirely in line with the normal abundance patterns
exhibited by most intervening DLAs. This is readily
appreciated from inspection of Figures~\ref{fig:plot_Zn_abund}
and \ref{fig:plot_ZnCr}, which compare the metallicity
(as measured by [Zn/H]) and the degree of dust depletion
(as indicated by the [Cr/Zn] ratio) of the PDLA with
the large compilation of analogous measurements 
assembled by \cite{akerman05}. It can be seen that,
with [Zn/H]\,$ = - 0.71$, the metallicity of the PDLA is
$2.7$ times higher than the median 
[Zn/H]\,$ = - 1.14$ of $\sim 50$ DLAs 
at $z_{\rm abs} = 2 - 3$, but still well within the range 
of metallicities encountered in intervening
DLAs. Similarly, the depletion of Cr relative to Zn by 
a factor of $\sim 2$ is typical of DLAs with metallicities
of $\sim 1/5$ solar.

We return to this
discussion in \S\ref{sec:discussion} below, after considering
the importance of: (a) partial coverage of the 
QSO continuum, and (b) the QSO
ionizing radiation, for the abundance determinations 
we have just presented. 
For readers who may not be interested 
in the details of these calculations, and wish
to proceed straight to \S\ref{sec:CII*},
the main conclusion of the following
two sections is that neither effect 
is likely to introduce significant corrections
to the abundances in Table~\ref{tab:abundances} and Figure~\ref{fig:plot_abundances}.

\section{Partial Coverage}\label{sec:partial_coverage}

In the course of absorption line fitting
(\S\ref{subsec:ion_col_densities}) we found evidence that component
C3, which is redshifted relative to $z_{\rm sys}$, does not fully
cover the continuum source. We now consider that evidence in more
detail.  If we examine the PDLA absorption lines in
Figures~\ref{fig:plot_normalised_vpfit.0}--\ref{fig:plot_normalised_vpfit.6}
and most directly in the top panels of
Figure~\ref{fig:plot_velocity_structure}, it is obvious that the
component with the highest optical depth in the species that are
dominant in H\,{\sc i} regions---component C1 centered at $v =
-162$\kms\ relative to $z_{\rm sys} = 2.57393$ (see
Table~\ref{tab:abs_comp})---has no residual flux in the core of the
strongest absorption lines. The damped \lya\ line itself shows no
discernible flux leaking in the line core within the accuracy with
which the background level can be determined; clearly this component
must cover completely not only the continuum emitting region but also
the more extended broad emission line region.  On the other hand, the
redshifted absorption at velocities between $v = +150$ and $+600$\kms\
does not reach below $\sim 0.2$ of the continuum flux in any of the
transitions covered by the 45.5\kms\ resolution ESI spectrum, nor
could it be fitted satisfactorily with VPFIT.

\begin{figure}
\begin{center}
\centerline{\includegraphics[width=0.9\columnwidth]{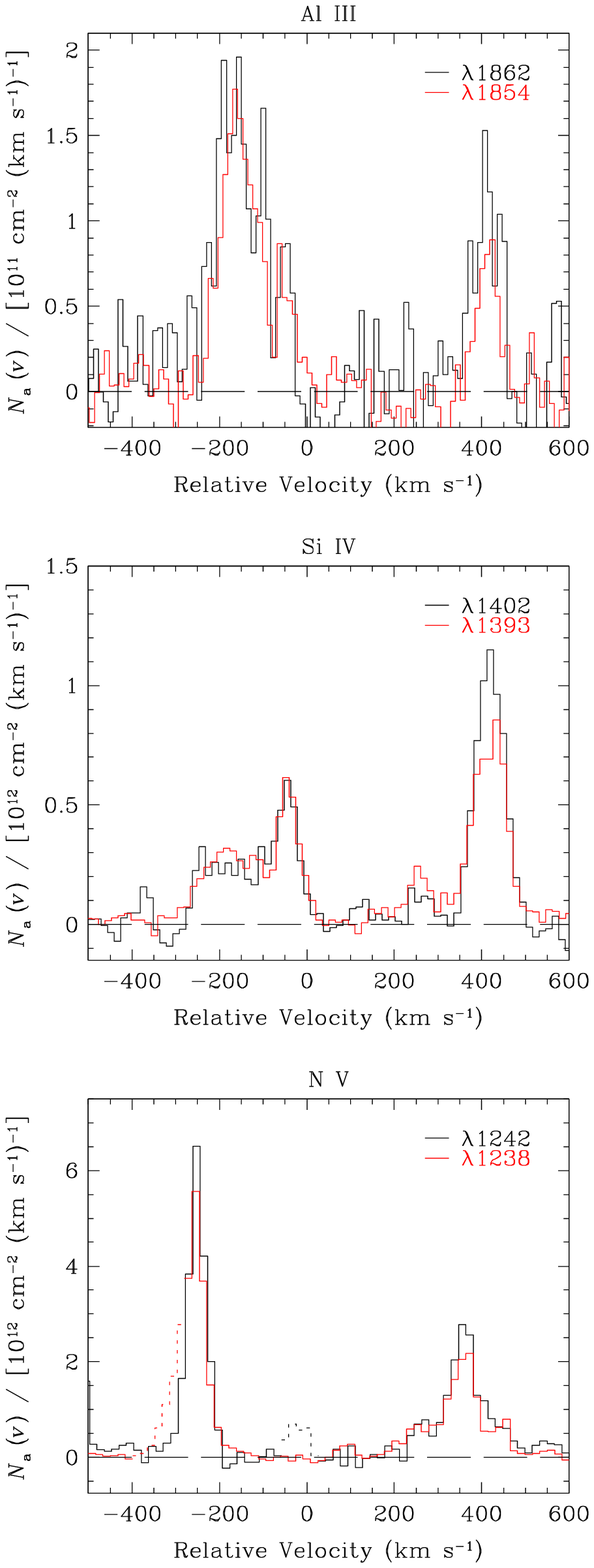}}
\caption{Apparent optical depth analysis of moderately and highly
ionized species. Apparent column densities per unit velocity are
displayed as a function of velocity (relative to $z_{\rm sys} =
2.57393$); blending with lines from other species is flagged by a
dotted line type. The transitions shown in the legends are in
increasing order of $f \lambda$ from the top.  The higher values of
$N_a(v)$ derived for the weaker, longer wavelength, members of the
Al\,{\sc iii}, Si\,{\sc iv} and N\,{\sc v} doublets in the redshifted
component of the absorption are suggestive of partial coverage of the
QSO continuum by the absorbing gas.\label{fig:plot_odm_high}}
\end{center}
\end{figure}

\begin{figure}
\begin{center}
\centerline{\includegraphics[width=0.9\columnwidth]{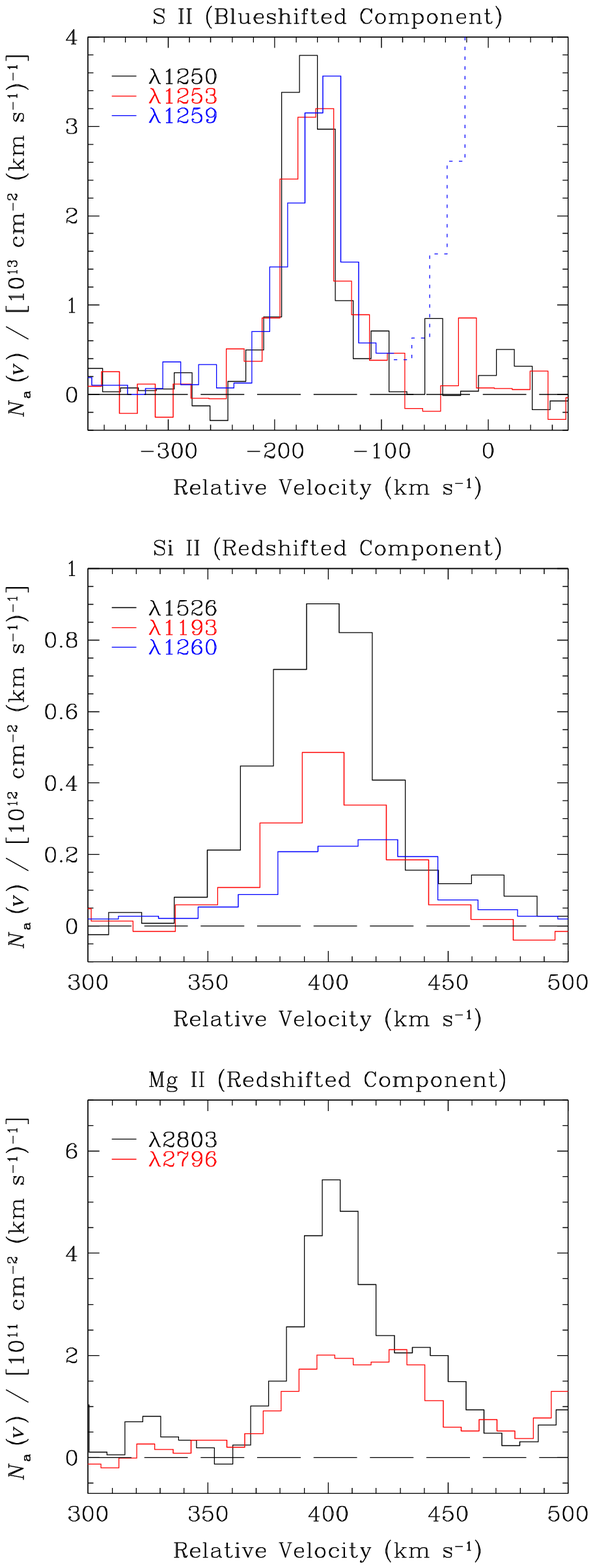}}
\caption{Apparent optical depth analysis of selected low ions.  The
transitions shown in the legends are in increasing order of $f
\lambda$ from the top.  The blueshifted absorption component is shown
for the S\,{\sc ii} triplet (top panel), while the redshifted
component is shown for three Si\,{\sc ii} lines (middle panel) and for
the Mg\,{\sc ii} doublet (bottom panel). Blending with lines from
other species is flagged by a dotted line
type.\label{fig:plot_odm_low}}
\end{center}
\end{figure}

We can address this issue more quantitatively by examining
the run of apparent optical depth with velocity
for transitions that arise from
the same ground state of a given ion.
The apparent column density of an ion 
in each velocity bin, $N_a(v)$ in units 
of cm$^{-2}$~(km~s$^{-1}$)$^{-1}$, 
is related to the apparent optical depth in 
that bin $\tau_a(v)$ by the expression
\begin{equation}
N_a(v) ~=~ \frac{\tau_a(v)}{f\lambda} \times \frac{m_e c}{\pi e^2} 
~=~  \frac{\tau_a(v)}{f \lambda (A)} \times 3.768 \times 10^{14}
\label{eq:tau1}
\end{equation}
where the symbols $f$, $\lambda$, $c$, $e$ and $m_e$ 
have their usual meanings.
The apparent optical depth can be deduced directly 
from the observed intensity
in the line at velocity $v$, $I_{\rm obs}(v)$, by
\begin{equation}
\tau_a(v) ~=~ -\ln~ [I_{\rm obs}(v)/I_0(v)]
\label{eq:tau2}
\end{equation}
where $I_0(v)$ is the intensity in the continuum.
With the assumption of negligible smearing of the 
intrinsic line profile by the 
instrumental broadening function, we can approximate
\begin{equation}
\tau_a(v) \approx \tau(v),
\label{eq:tau3}
\end{equation}
where $\tau(v)$ is the true optical depth.
The run of $N_a(v)$ with $v$ should be the same, within the errors
in $I_{\rm obs}(v)$, for all lines arising from the same atomic energy level,
provided there are no unresolved saturated components to the
absorption lines and the covering factor is unity \citep{savage91}.
The assumption in equation~(\ref{eq:tau3}) is probably not justified
in our case: observations of damped Lyman alpha systems
at $\sim 6$\kms\ resolution show that absorption components
much narrower than the 45.5\kms\ resolution of the ESI data 
presented here are not unusual \citep{wolfe05}.
Even so, lack of agreement in $N_a(v)$ between different transitions
would still be an indication of saturation or partial
coverage\footnote{In the following discussion, we use the nomenclature
{\em saturation\/} to refer to line saturation arising in the
`typical' scenario that the intervening gas has a uniform covering
factor ($f=1$), while {\em partial coverage\/} is adopted for cases
where there is a non-uniform covering factor. Note that the latter
definition does not exclude the possibility that the absorption lines
also suffer from saturation.}, although the opposite may not be true.

Among the ion transitions covered in our spectrum of Q2343--BX415,
there are three doublets whose apparent optical depths can be 
compared over the full velocity range of the absorption (or nearly so):
Al\,{\sc iii}, Si\,{\sc iv} and N\,{\sc v}; in all other cases
either the main absorption component is saturated, or the redshifted
absorption is too weak, or one of the lines is seriously affected by blending.
Values of $N_a(v)$ for the three doublets are 
shown in Figure~\ref{fig:plot_odm_high}. 
For Al\,{\sc iii} and Si\,{\sc iv} we see  good
agreement, within the noise, in the blueshifted absorption
(components C1 and C2), but not in the redshifted absorption
(C3). The situation is less clear-cut in N\,{\sc v}, 
where there may be disagreement in the values of 
$N_a(v)$ between the two members of the doublet
in both the blueshifted and redshifted components.
As noted earlier, N\,{\sc v} 
is displaced in velocity relative to Al\,{\sc iii}, Si\,{\sc iv} and
C\,{\sc iv}, attesting to the complex nature of the absorbing gas.

When it comes to species that are the dominant ions in
H\,{\sc i} regions and on which the abundance analysis
in \S\ref{subsec:abundances} was based, we can only 
perform the optical depth consistency check on individual
components, rather than on the full range of the absorption,
for the reasons given above. Even so, the illustrative
comparisons reproduced in Figure~\ref{fig:plot_odm_low}
are instructive. 
Component C1 is well behaved in S\,{\sc ii} (top),
while C3 again shows evidence of saturation or partial coverage
in Si\,{\sc ii} (middle) and Mg\,{\sc ii} (bottom).

We consider that the discrepancies in the $N_a(v)$ profiles of C3 most
likely arise from partial coverage, rather than simply saturation, for
the following reasons. First we consider the higher ionization
Al\,{\sc iii} lines, for which we can directly compare the profiles of
both blueshifted and redshifted absorption. If the discrepancy
observed in the {\em weaker\/} redshifted component were due to
saturation, the stronger blueshifted components would also be
saturated and should display discrepancies in the line profiles, in
contrast with what is observed. It therefore appears that the third,
redshifted, component of the moderately ionized gas suffers from
partial coverage. Of course we cannot be certain that the lower
ionization gas is spatially coincident with the higher ionization
gas. Nevertheless it seems unlikely that the discrepancies in $N_a(v)$
observed for the low ionization gas are due to saturation as no
discrepancy is seen for the main component of \siiva\ (see
Fig~\ref{fig:plot_odm_high}), which has a comparable optical depth to
the third components of some of the Si\,{\sc ii} and Mg\,{\sc ii}
lines.

To summarize, it appears that the geometrical 
distribution of the blueshifted
and redshifted absorptions are different,
with the former totally covering the QSO continuum 
and broad line emitting regions, while the latter
only does so partially or inhomogeneously.
If a significant fraction of the neutral gas resided in the
redshifted component C3, the 
element abundances we deduced in \S\ref{subsec:abundances}
may have been underestimated. Fortunately, C3
accounts for negligible amounts 
of the ions that are dominant in H\,{\sc i} regions.
Based on the lines with the lowest values of 
$f \lambda$ in which C3 is seen (these are the lines
where the effect of partial coverage on the derivation
of the column density is minimized), we estimate
from the optical depth and line fitting analyses that
the fraction of $N$(Mg\,{\sc ii}), $N$(Si\,{\sc ii}), and 
$N$(Fe\,{\sc ii}) contributed by C3 is in all three
cases $\lesssim 1$\%.

\section{Ionization Corrections}\label{sec:ionization_corrections}

In \S\ref{subsec:abundances} we derived element abundances
by relating the column density of each element 
{\em in its dominant ionization state in neutral gas\/} 
to that of neutral hydrogen. 
Referring to equation~(\ref{eq:ionization_assumption}),
this assumption neglects the possibility that the line
of sight may include (partly) ionized gas where
$N$(X$^i$)/$N$(H\,{\sc i})\,$\neq N$(X)/$N$(H).
For example, a fraction of the column density $N$(X$^i$) 
may arise in ionised gas, leading to an \emph{over}estimate 
of the true abundance of element X; or an element may
be over-ionized compared with H, resulting in an 
\emph{under}estimate of its abundance.
 
This has been a potential problem for interstellar abundance
studies since their very beginning \citep[\eg][]{steigman75}.
\citet{vladilo01} considered the issue in depth for intervening
DLAs exposed to the metagalactic ionizing background
and stellar UV radiation from their host galaxies.  
The principal conclusion of that work was that for most DLAs the 
contamination from H\,{\sc ii} regions leads to abundance
corrections that are generally less than $0.1-0.2$\,dex,
smaller than, or comparable to, other sources of error.

In proximate DLAs, however, the corrections may well be larger,
depending on the impact of the QSO's own radiation on the ionization
structure of the gas.  We examined this possibility closely by running
a series of photoionization models with the software package CLOUDY
\citep[v05.07.06;][]{ferland98,ferland00}\footnote{http://www.nublado.org/}.
The PDLA is treated as a slab of gas of constant density upon which a
variety of radiation fields are impinging.  For Q2343--BX415 itself we
adopted an AGN ionizing continuum with a power-law slope
$\alpha_\nu=-0.44$ (see \S\ref{sec:spectrum_bx415}); in addition we
included the metagalactic ionizing background \citep{haardt01} and the
cosmic microwave background, both at $z=2.572$, as well as cosmic ray
ionization at a level comparable to that in our own Galaxy.  We
adjusted the CLOUDY element abundance scale to reflect the abundance
pattern determined in \S\ref{subsec:abundances}, which is a mixture of
genuine underabundances and dust depletions, and added dust grains in
the right proportions to reflect the latter.

We ran a grid of CLOUDY models by varying two parameters:
the gas density ($\log [n{\rm (H)}/{\rm cm}^{-3}] = -3$ to $+3$) 
and the ionization parameter ($\log U = -5$ to $+1$)
\footnote{The ionization parameter $U$ used by CLOUDY is the
dimensionless ratio between the hydrogen ionizing photon density and
the gas density. It can be expressed as 
$U=\phi{\rm (H)}/{n{\rm (H)}c}$, 
where $\phi$(H) is the surface flux of ionizing
photons, $n$(H) is the total hydrogen density 
and $c$ is the speed of light.}.
For each combination of $n{\rm (H)}$ and $U$ 
we stopped the calculation when the measured
column density of neutral gas, 
$N$(H\,{\sc i})\,$ = 9.5 \times 10^{20}$\,cm$^{-2}$
(\S\ref{subsec:p-dla}), was reached and computed,
for each element observed,  
the difference between the input abundance and that
which would be deduced by applying the assumption of 
equation~(\ref{eq:ionization_assumption}). 
This defines an ionization correction
\begin{equation}
{\rm IC(X)} = \log \left[ \frac{N{\rm (X)}}{N({\rm H})}\right]_{\rm
intrinsic} - \log \left[ \frac{N{\rm (X^{\mathit i})}}{N({\rm
H{\scriptstyle~I}})}\right]_{\rm computed}
\label{eq:IC(X)}
\end{equation}
which is negative when the abundance of an element has been
overestimated by neglecting the ionized gas.

\begin{figure}
\centerline{\includegraphics[width=0.9\columnwidth]{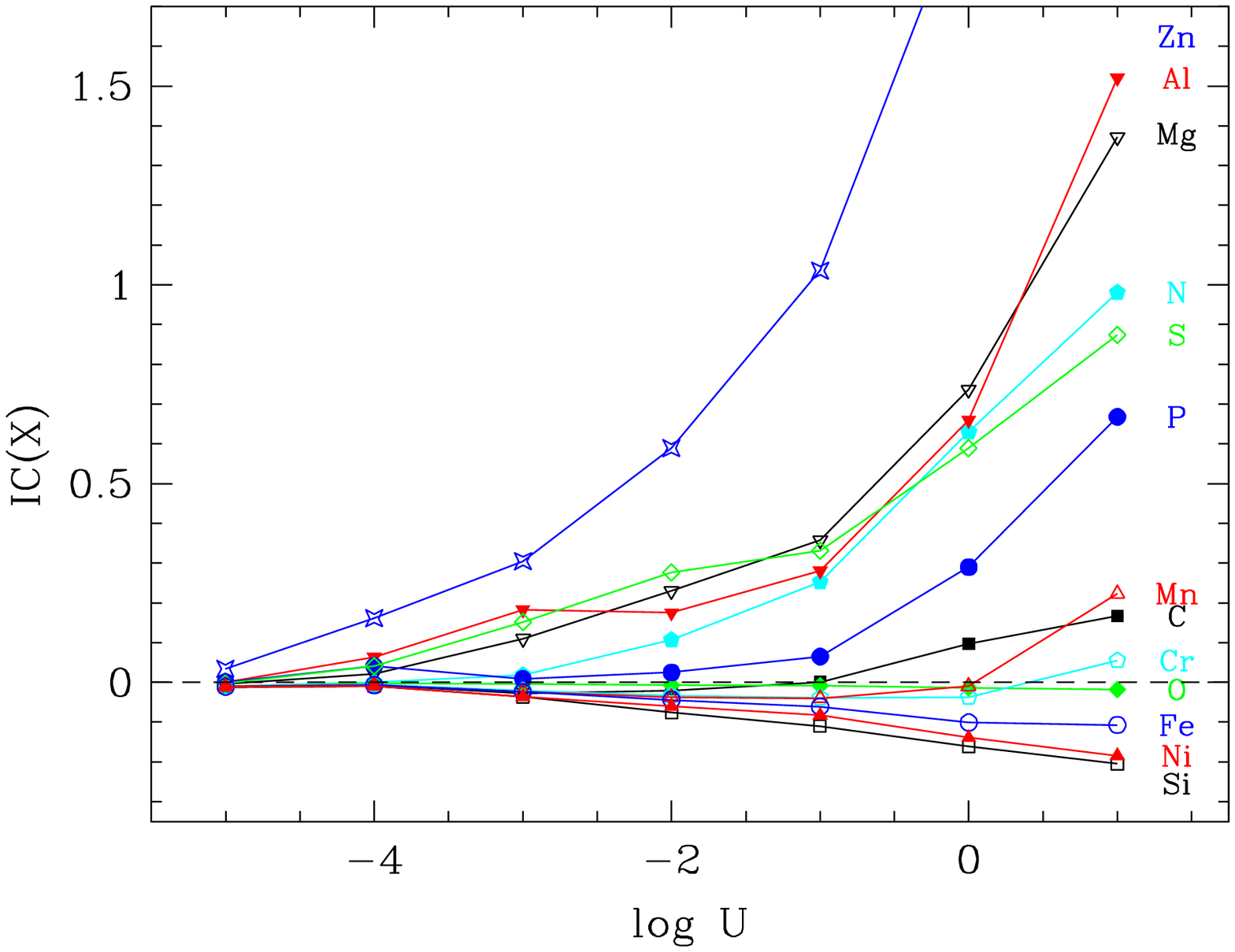}}
\caption{Logarithmic ionization corrections applicable
to our abundance estimates in the PDLA 
(see equation~\ref{eq:IC(X)}) plotted
as a function of the ionization parameter $U$,
for a constant density model with $n$(H)\,$= 10$\,cm$^{-3}$.
The corrections were computed using the photoionization code CLOUDY.\\\label{fig:plot_ic}}
\end{figure}

Figure~\ref{fig:plot_ic} shows the magnitude of such ionization
corrections for different elements as a function of $\log U$ 
for a representative case with $\log [n{\rm (H)}/{\rm cm}^{-3}] = +1$; 
the results for other values of gas density are qualitatively similar.
Clearly, the impact of the ionization corrections on our estimates
of element abundances depends sensitively on the value of the 
ionization parameter, at least for some elements. In order to make
progress it is therefore necessary to pin down, or at least constrain,
the value of $U$.

\begin{figure}
\centerline{\includegraphics[width=0.9\columnwidth]{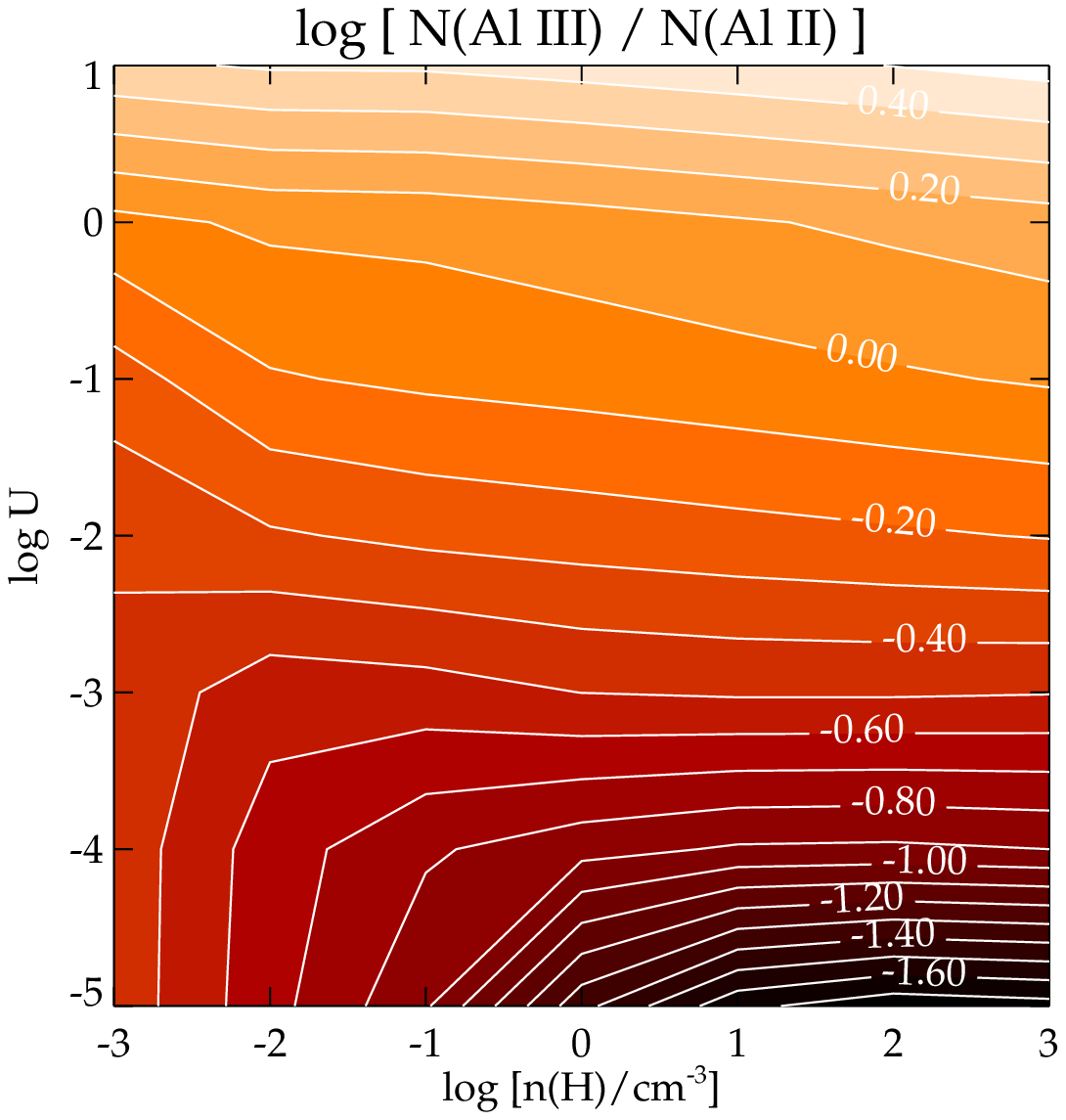}}
\caption{Values of  
$\log [N$(Al\,{\sc iii})/$N$(Al\,{\sc ii})] 
computed by CLOUDY for a range of 
ionization parameters $U$ and hydrogen
densities $n{\rm (H)}$.\\\label{fig:plot_al_ratio}}
\end{figure}

This can be achieved by considering the ratios of different
ionization stages of the same element; as found by \citet{vladilo01},
the ratio of Al\,{\sc iii} and Al\,{\sc ii} is of particular diagnostic
value in this context. 
This is illustrated in Figure~\ref{fig:plot_al_ratio} where the
values of $\log$[$N$(Al\,{\sc iii})/$N$(Al\,{\sc ii})]
computed by CLOUDY are color-coded as a function of 
both $U$ and $n$(H);
clearly the major dependence is on $U$ 
(in the sense that the ratio increases with the ionization parameter)
rather than $n$(H).  

Referring to Figures~\ref{fig:plot_normalised_vpfit.2} and
\ref{fig:plot_normalised_vpfit.3} respectively, it can be seen that
while Al\,{\sc ii}~$\lambda 1670$ is saturated, the lines of the
Al\,{\sc iii} doublet at $\lambda\lambda 1854, 1862$ are weak.
Profile fitting with VPFIT (shown in red in the Figures), gives
$N$(Al\,{\sc iii})\,$= 2.1 \times 10^{13}$\,cm$^{-2}$ and $N$(Al\,{\sc
ii})\,$\geq 3.7 \times 10^{14}$\,cm$^{-2}$ for components C1 and C2
\footnote{The first (second) component of Al\,{\sc iii} was fitted
with a column density $N = 1.7\times 10^{13}$\,cm$^{-2}$ ($4.0\times
10^{12}$\,cm$^{-2}$), Doppler parameter $b=48.9$\kms\ ($26.5$\kms),
and redshift $z=2.57203$ ($2.57330$).}, from which we deduce:
\begin{equation}
\left[ \frac{N({\rm Al\,{\scriptstyle III}})}
{N({\rm Al\,{\scriptstyle II}})}\right] _{\rm observed}
\leq 10^{-1.25}.
\end{equation}
From Figure~\ref{fig:plot_al_ratio} we find that such low ratios 
can only be achieved (in the broad parameter space that we 
explored) if $\log U \lesssim -4.3$. This in turn implies small
ionization corrections: ${\rm IC(X)} < 0.1$ for all elements
apart from Zn (Figure~\ref{fig:plot_ic}).\footnote{It has been 
suspected for some time that there may be problems with the  
ionization balance computed by CLOUDY for Zn
\citep[\eg][]{howk99, vladilo01}, presumably as a result of incorrect 
atomic data relevant to the calculations of the 
ionization and recombination rates.}

It is remarkable that the Al\,{\sc iii}/Al\,{\sc ii} ratio is so low,
given the proximity of Q2343--BX415 to the DLA.
The ratio we measured here is, if anything, towards the
lower end of the distribution of values found in 
conventional DLAs at $z_{\rm abs} \ll z_{\rm em}$
\citep[see Fig.~2 of][]{vladilo01}. 
Evidently, the neutral gas in this PDLA is shielded
from the additional ionizing radiation provided by
the nearby quasar. Even though there is absorption
by a range of ionization stages over the same velocity
intervals (see Figure~\ref{fig:plot_velocity_structure}),
the ionized and neutral gas presumably occur in 
physically distinct regions.
For the purposes of the present discussion,
we conclude that the element abundance analysis
presented in \S\ref{subsec:abundances} is unlikely 
to be in serious error as a result of the proximity
of the DLA to the background QSO.

\section{THE DISTANCE OF THE PDLA FROM Q2343--BX415}\label{sec:CII*}

Among the absorption lines we detect in the PDLA is 
C\,{\sc ii}$^{\ast}$~$\lambda 1335.7032$, which arises
from the excited fine-structure $^2P_{3/2}$ level of the
ground state of singly ionized carbon. The absorption
line can be seen clearly in the middle panel of 
Figure~\ref{fig:plot_normalised_vpfit.1}. 
Attention has focused on this line in DLAs
since \cite{wolfe03a} pointed out that it could
be used to infer the strength of the radiation field
to which the gas is exposed. Briefly, the argument is 
based on the fact that the spontaneous downward transition
to the $^2P_{1/2}$ ground-state (from which the stronger
C\,{\sc ii}~$\lambda 1334.5323$ absorption line arises)
at 158\,$\mu$m is likely to be the major coolant in 
H\,{\sc i} regions. In thermal equilibrium,  
the cooling rate is expected to balance the heating rate;
since the latter is primarily due to photoelectric emission from
dust grains, the cooling rate inferred from the ratio
$N$(C\,{\sc ii}$^{\ast}$)/$N$(H\,{\sc i}) can be used
to infer the intensity of the radiation field responsible
for heating the gas. In PDLAs this line is of particular
interest, as a means of gauging the contribution of the
QSO to the radiation field impinging on the DLA, and
thereby obtaining an indication of their physical separation. 

We begin by working out the spontaneous 
[C\,{\sc ii}]~158\,$\mu$m emission rate per hydrogen atom:
\begin{equation}
l_{\rm c} = 
\frac{N({\rm C{\scriptstyle\,II}^{\ast}})}{N({\rm H{\scriptstyle\,I}})}
\times h \nu_{\rm ul} \, A_{\rm ul}
~=~ 
\frac{N({\rm C{\scriptstyle\,II}^{\ast}})}{N({\rm H{\scriptstyle\,I}})}
\times 3.0 \times 10^{-20}
~{\rm ergs~s}^{-1}~{\rm H~atom}^{-1}
\label{eq:cii*1}
\end{equation}
where $h \nu_{\rm ul}$ is the energy of a 158\,$\mu$m photon and
$A_{\rm ul}$ is the Einstein coefficient for spontaneous photon
decay. From our VPFIT analysis\footnote{Note that we did not constrain
the redshift and Doppler parameter of the C\,{\sc ii}$^{\ast}$ gas
with those values given in Table~\ref{tab:abs_comp} due to the
different excitation levels of the gas; VPFIT fitted a single
component at $z=2.57188$ with $b=23.0$\kms.}, we deduce $N$(C\,{\sc
ii}$^{\ast}$)\,$=1.33 \times 10^{14}$\,cm$^{-2}$; combining this value
with $N$(H\,{\sc i})\,$= 9.5 \times 10^{20}$\,cm$^{-2}$
(\S\ref{subsec:p-dla}), we deduce from equation~(\ref{eq:cii*1}):
\begin{equation}
\log (l_{\rm c}/ {\rm ergs~s}^{-1}~{\rm H~atom}^{-1}) = -26.38
\label{eq:cii*2}
\end{equation}

\begin{figure}
\centerline{\includegraphics[width=0.925\columnwidth]{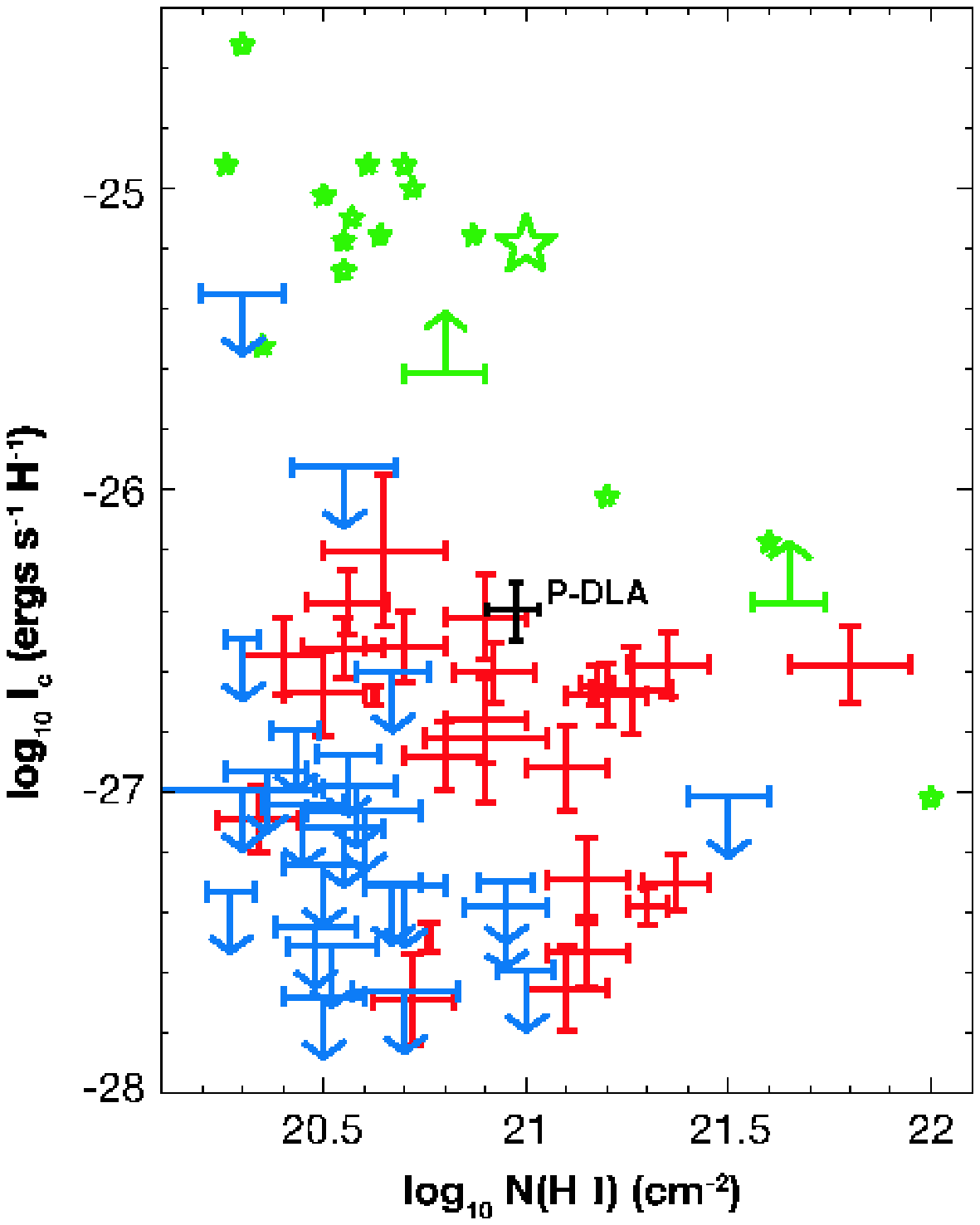}}
\caption{Compilation of available measurements of the [C\,{\sc
ii}]~158\,$\mu$m photon emission rate per H atom, $l_{\rm c}$, in DLAs
\citep[adapted from][]{wolfe05}. Error bars are used for DLAs where
$N$(C\,{\sc ii}$^{\ast}$) has been measured ({\em red in the
electronic edition}), while downward arrows ({\em blue in the
electronic edition}) are $2 \sigma$ upper limits. Two lower limits
(also $2 \sigma$) are shown by upward arrows ({\em green in the
electronic edition}) and the value deduced here for the PDLA in
Q2343--BX415 is indicated in black.  Small stars are values of $l_{\rm
c}$ measured along different sight-lines through the interstellar
medium of the Milky Way; the large star is the value averaged over the
disk of our Galaxy. The higher dust content of the Milky Way ISM is
thought to be the main reason for the higher cooling (and heating)
rates compared to most DLAs.\label{fig:plot_cooling_rate}}
\end{figure}

This value is not dissimilar to those measured in some
intervening DLAs. 
Figure~\ref{fig:plot_cooling_rate} shows the most recent compilation
by \cite{wolfe05} of values of $l_{\rm c}$ measured in DLAs.
While in about half the cases $\log l_{\rm c} \lesssim -27$,
values as high as that deduced here
for the PDLA in Q2343--BX415 
have been reported in a few cases.

\cite{wolfe03a} developed the radiation transfer formalism
required to interpret the values of $l_{\rm c}$ in terms of $J_{\nu}$,
the far-UV radiation intensity responsible for the heating rate
of the gas. The main conclusion of their analysis,
also discussed by \cite{wolfe05}, is that while 
the DLAs where C\,{\sc ii}$^{\ast}$\,$\lambda 1335.7$ is undetected
(the upper limits in Figure~\ref{fig:plot_cooling_rate})
are consistent with the value of $J_{\nu}$ due to the 
integrated extragalactic background, 
$J_{\nu}^{\rm tot} = J_{\nu}^{\rm bg}$,
those with higher values of $l_{\rm c}$ 
require an additional contribution $\Delta J_{\nu}$,
so that $J_{\nu}^{\rm tot} = J_{\nu}^{\rm bg} + \Delta J_{\nu}$.
\cite{wolfe03a} identify this additional component $\Delta J_{\nu}$
as the UV radiation from young stars within the galaxies hosting 
the DLAs and use this method to estimate the star formation
rate of such systems.

We can follow the same line of reasoning and deduce the
maximum contribution of Q2343--BX415 to the radiation
field to which the PDLA is exposed, under the 
assumption of negligible contribution from \emph{in situ} 
star formation. Comparison of the value of $\Delta J_{\nu}$
appropriate to the PDLA with the known luminosity of the 
quasar, would then provide a lower limit to the distance
between Q2343--BX415 and the PDLA---any 
local stellar contribution to $\Delta J_{\nu}$ would place
the absorber further away from the quasar.

Applying the radiation transfer code of \cite{wolfe03a}\footnote{ The
calculation was kindly performed for us by Dr. A.~M. Wolfe.} with the
parameters appropriate to the PDLA---redshift, metallicity and
dust-to-gas ratio (deduced from the observed [Fe/Zn] ratio, see
\S\ref{subsec:abundances}), we find that our measured $\log l_{\rm c}
= -26.38$ implies $\Delta J_{\nu} = 2.0 \times
10^{-18}$\,erg~s$^{-1}$~cm$^{-2}$~Hz$^{-1}$~sr$^{-1}$ if the C\,{\sc
ii}$^{\ast}$ absorption arises in a `Cold Neutral Medium' (CNM), and
$\Delta J_{\nu} = 4.4 \times 10^{-17}$ (in the same units) if it
arises in a `Warm Neutral Medium' (WNM).  The former solution is the
one generally adopted for DLAs (see Wolfe et al. 2003 and Wolfe et
al. 2005 for further details).

These values of $\Delta J_{\nu}$ correspond to the excess
over the \cite{haardt01} extragalactic background at a
wavelength of 1500\,\AA. The distance from Q2343--BX415
at which the QSO contributes this excess is then given by:
\begin{equation}
d = \frac{1}{4 \pi} \, \sqrt{\frac{L_{\nu}}{J_{\nu}}}
\label{eq:cii*3 }
\end{equation}
where $L_{\nu}$ is the QSO luminosity at 1500\,\AA. 
We calculate this last quantity from the measured
$G = 20.32$ broad-band magnitude, extrapolating the 
QSO continuum to $\lambda_0 = 1500$\,\AA, and
after small corrections to account for filter transmission curves, 
emission and absorption lines. We find 
$L_{1500}^{\rm Q2343-BX415} = 4.1 \times 10^{30}$\,erg~s$^{-1}$~Hz$^{-1}$
and therefore $d \geq 37$\,kpc for the 
CNM solution. The WNM solution, if applicable,
would lower the limit to $d \geq 8$\,kpc.

\section{DISCUSSION}\label{sec:discussion}

The main results of the work presented in this paper are as
follows.

(1) We have obtained a high signal-to-noise ratio, medium resolution,
spectrum of the faint QSO Q2343--BX415 motivated by the presence
of a strong DLA coincident in redshift with the QSO. Our intention
was to investigate gas in the vicinity of the quasar; in particular, if the DLA
arises in the host galaxy of Q2343--BX415 it would allow a more 
comprehensive characterization of its chemical composition 
than is normally possible with other techniques that are based either on high 
ionization absorption systems at $z_{\rm abs} \approx z_{\rm em}$
or on the analysis of  emission lines from the AGN.
Unlike previous studies of other proximate systems,
we are aided by the precise knowledge of both absorption and
emission redshifts, the former from metal lines in the DLA 
and the latter from narrow emission lines in the rest-frame optical
and UV spectrum of Q2343--BX415.

(2) We find that metal lines associated with the DLA
consist of two sets of absorption components. One component is 
apparently moving towards the QSO, being redshifted by 
$\sim 150 - 600$\kms\ relative to the systemic redshift 
$z_{\rm sys} = 2.57393$. This component is prominent in 
high ionization lines and does not fully cover the continuum 
source, suggesting that it is physically close to the 
active nucleus of the galaxy. To our knowledge, this is
the first time that gas falling onto the nucleus
has been seen in absorption.

(3) The second set of absorption components is blueshifted,
with velocities ranging from $+50$\kms\ to $-325$\kms\ relative to 
$z_{\rm sys}$; the neutral gas has maximum optical depth 
in a component at $v = - 162$\kms. 
These velocities are of the same order of magnitude 
as those often encountered within star-forming galaxies at
$z = 2 - 3$; in particular the shift of $-162$\kms\ of the bulk of the neutral
gas is the same as the mean offset between UV absorption lines
and H$\alpha$ emission measured in nearly one hundred
BX galaxies at $z \simeq 2.0$--2.6 by Steidel et al. (in preparation). 
Such velocity
offsets are normally interpreted as evidence for galactic-scale
outflows in star-forming galaxies, suggesting that in the
PDLA we may be seeing the outflowing interstellar medium 
of the host galaxy of Q2343--BX415.

(4) The blueshifted absorption includes gas over a
wide range of ionization, from neutral species 
to C\,{\sc iv} and N\,{\sc v}. The latter in particular 
is indicative of a strong non-thermal ionizing source,
presumably due to the combination of the close proximity
of \emph{two} QSOs: Q2343--BX415 and Q2343+1232,
which is located only $\sim 690$\,kpc away from Q2343--BX415.
However, the gas in the DLA itself appears to be 
shielded from the Lyman continuum radiation 
from the two QSOs and from the metagalactic background:
the low $N$(Al\,{\sc iii})/$N$(Al\,{\sc ii}) ratio
is indicative of a low ionization parameter for the 
neutral gas. This in turn implies that it is possible to deduce
element abundances without significant corrections 
for unobserved ion stages.

(5) The presence of C\,{\sc ii}$^{\ast}$ in the PDLA
gives a measure of the UV radiation field longward
of the Lyman continuum, under the assumption of
equilibrium between the cooling rate via 
[C\,{\sc ii}]~$\lambda 158\,\mu$m emission
and photoelectric heating from grains. 
The value of $J_{1500}$ we deduce, adopting
the radiation transfer formalism of \cite{wolfe03a},
is $\sim 50$ to $\sim 1000$ times greater than the 
metagalactic radiation field of \cite{haardt01},
depending on the temperature of the gas.
If Q2343--BX415 is the main source of this intense
radiation field, the PDLA must be 
located only a few tens of kpc from the QSO,
more specifically at $\sim 8$\,kpc or $\sim 37$\,kpc depending on whether
the C\,{\sc ii}$^{\ast}$ absorption arises in a
`Cold' or a `Warm Neutral Medium'.

(6) On the other hand, in the compilation by \cite{wolfe05} there are
other intervening DLAs where the rate of grain photoelectric heating
is as high as that measured here. Those authors attribute the excess
radiation (over the metagalactic background) to OB stars in the host
galaxies of the DLAs; any such contribution would increase the
inferred distance of the PDLA from Q2343--BX415.  While this remains a
possibility, we note that: (a) a stellar ionizing spectrum would not
account for the presence of strong N\,{\sc v} in the blueshifted
absorption, which would then have to arise in a physically distinct
region from the DLA \citep[see for example][]{fox07}; and (b) as
discussed in Appendix~\ref{sec:intervening_systems}, our deep imaging
and follow-up spectroscopy of the field of Q2343--BX415 has not yet
produced a plausible galaxy candidate (other than the host galaxy of
the QSO) for the PDLA.

(7) The overall metallicity of the DLA and its detailed chemical
composition in 14 elements of the periodic table are similar to those
encountered in the normal DLA population. With a metallicity of $\sim
1/5$ solar, the PDLA is a factor of $\sim 2 - 3$ more chemically
enriched than the median at its redshift, but there is no evidence of
the super-solar metallicities that have been claimed for associated,
high ionization systems \citep[\eg][]{petitjean94}.  No peculiarities
were found in the pattern of relative abundances of the different
elements sampled, most of the departures from solar relative
abundances being compatible with depletions of refractory elements
onto dust grains. Interestingly, there is a remarkable similarity, in
both the degree of metal enrichment and the relative abundances,
between this PDLA and the other well studied PDLA to date, the $z_{\rm
  abs} = 2.8110$ system towards Q0528$-$2505
\citep{lu96,ledoux06}.\footnote{For this comparison, recent revisions to the
  solar abundance scale and the $f$-values of relevant atomic
  transitions must be taken into consideration.}

However, the chemical data are not of clear diagnostic value regarding
the origin of the PDLA.  Its overall metallicity of $\sim 1/5$ solar
is intermediate between the values typical of intervening DLAs
\citep[$\langle Z_{\rm DLA} \rangle \simeq 1/15\, Z_{\odot}$,
][]{akerman05} and of star-forming galaxies \citep[$ Z_{\rm LBG}
\simeq 1/3 - 1 \, Z_{\odot}$, ][]{erb06a, pettini01} at redshifts $z =
2 -3$.  Its detailed abundance pattern is consistent with either
interpretation \citep{wolfe05, pettini02}.  The finding that the
metallicity is a factor of $\sim 2 - 3$ lower than that of most of the
BX galaxies studied by \cite{erb06a} may be an indication that the
host galaxy of Q2343--BX415, if that is what we are detecting, is
intrinsically (\ie\ when its nucleus is not in an active phase)
fainter than the ${\cal R} = 25.5$ limit of the BX survey.
Alternatively, there may well be metallicity gradients between the
central, star-forming regions of UV-bright galaxies (to which the
measures of Erb et al. 2006a refer) and outer regions where the
interstellar medium is predominantly neutral.  However, such an effect
is not seen in MS~1512-cB58, the only galaxy at $z = 2 - 3$ where such
a comparison has been made up to now \citep[][]{pettini02, rix04}.

In summary, our detailed observations of the rest-frame
UV spectrum of Q2343--BX415 provide us with several
clues as to the origin of the proximate DLA, but it
is not yet possible to fit all of them into an entirely
self-consistent interpretation.
On the one hand, the PDLA could result 
from the chance alignment of a foreground galaxy and Q2343--BX415. 
When compared with the bulk of the
DLA population, 
there is nothing particularly unusual in its kinematics, 
metal enrichment, abundance pattern, 
and low degree of ionization reflected in the low
Al\,{\sc iii}/Al\,{\sc ii} ratio.
The relatively strong C\,{\sc ii}$^{\ast}$ absorption
could be due to \emph{in situ} star formation.
However, this picture requires \emph{two} velocity 
matches to be coincidental: not only between the redshifts
of the absorber and the quasar host, but also
between the neutral gas in the DLA and the high
ions---particularly N\,{\sc v}---which in this picture 
would have to located in a 
separate region from the neutral interstellar
medium of the DLA galaxy.

On the other hand, if the DLA is physically associated with 
Q2343--BX415, then our observations
indicate that the gas is outflowing from the host galaxy of the quasar
with velocities and other properties that are very much 
in line with those of the `superwinds' 
seen in most star-forming galaxies 
at comparable redshifts.
The placement of the neutral gas at a few
tens of kpc from the active nucleus
(from the C\,{\sc ii}$^{\ast}$ analysis) 
is suggestive of a shell of swept-up interstellar material.
Its relatively low metallicity 
(compared with those typically encountered in 
the star-forming regions of UV-bright galaxies) 
may be an indication that the host galaxy is 
fainter and of lower mass than most BX galaxies
studied so far---and yet still harbors an AGN.
However, the low Al\,{\sc iii}/Al\,{\sc ii} ratio
implies that the DLA must somehow be shielded
from the ionizing radiation from the active nucleus.

Regardless of the uncertain origin of the PDLA itself, 
we have probably detected gas associated with the
AGN in the (mostly ionized) redshifted components,
which have less than 100\% covering factor.
While it would be very interesting to examine 
the metallicity of these intrinsic absorbers---and
compare it to that of the emission line gas---both
determinations would be much more
model-dependent than the straightforward 
measurement of DLA abundances on which we 
have focused in this paper.
  
In closing, while we have not been able
to establish conclusively the origin of the PDLA,
observations as detailed as the ones 
presented here still emphasize the importance of
further studies of proximate DLAs as probes 
of the environments of quasars at early times. 
Comparison of their properties and physical conditions
with those of the `field' population of DLAs
will help us understand better the nature of 
DLAs at large, as well as add to our knowledge of
the galaxies, or groups of galaxies, that host AGN.
It is thus well worth targeting the fields of QSOs
with known PDLAs with the combined efforts of deep
galaxy surveys and high resolution absorption line
spectroscopy; the two techniques together have the 
potential to provide some of the most valuable 
empirical constraints on `AGN-driven feedback'
at high redshift.

\acknowledgments
It is a pleasure to acknowledge illuminating
discussions with Bob Carswell, Gary Ferland and Paul Hewett, and
valuable suggestions from the referee, which have improved this paper. We
are indebted to Art Wolfe who generously helped with the
interpretation of the C\,{\sc ii}$^{\ast}$ absorption.  Part of this
work was carried out during visits by Sam Rix and Art Wolfe to the
Institute of Astronomy, Cambridge, supported by the Institute's
visitors' grant. S. Rix also acknowledges the support of a PPARC
Postdoctoral Research Fellowship. We thank the staff of the Keck
Observatory for their expert assistance with the ESI and NIRSPEC
observations, and the Hawaiian people for the opportunity to observe
from Mauna Kea.  Without their hospitality, this work would not have
been possible.



\begin{appendix}

\section {Intervening Absorption Line Systems}\label{sec:intervening_systems}

In addition to the PDLA, we recognize 11 intervening absorption line
systems in our ESI spectrum of Q2343--BX415, at 
$z_{\rm abs} < z_{\rm sys}$ (see Table~\ref{tab:intervening_ews}).  
These systems range in redshift from $z_{\rm abs} = 1.2058$ to 2.4862, 
and are mostly high ionization C\,{\sc iv} systems, 
although we also have one Mg\,{\sc ii} system 
and a rich, mixed ionization, system 
at $z_{\rm abs}=2.1735$ in which we identify 17 absorption lines.
All the intervening absorption lines are 
labelled (S1 through to S11) in 
Figures~\ref{fig:plot_normalised_vpfit.0}--\ref{fig:plot_normalised_vpfit.6}
and, like those in the PDLA, were fitted with VPFIT to deduce
the values of column density listed in Table~\ref{tab:intervening_ews}.

\LongTables
\tabletypesize{\footnotesize}
\begin{deluxetable*}{clccccccll}
\tablewidth{0pt}
\tablecolumns{10}
\tablecaption{\textsc{Intervening Absorption Line Systems}}
\tablehead{
  \multicolumn{1}{c}{$\lambda_{\rm obs}$ \tablenotemark{a}}
& \multicolumn{1}{c}{Identification}
& \multicolumn{1}{c}{$z_{\rm abs}$}
& \multicolumn{1}{c}{$\Delta v$ \tablenotemark{b}}
& \multicolumn{1}{c}{$W_{\rm obs}$ \tablenotemark{c}}
& \multicolumn{1}{c}{$\sigma_{\rm obs}$}
& \multicolumn{1}{c}{$W_0$ \tablenotemark{d}}
& \multicolumn{1}{c}{$\sigma_0$}
& \multicolumn{1}{c}{$\log N$ \tablenotemark{e}}
& \multicolumn{1}{c}{Comments} \\
  \multicolumn{1}{c}{(\AA)}
& \colhead{ } 
& \colhead{ } 
& \multicolumn{1}{c}{(\kms)}
& \multicolumn{1}{c}{(\AA)}
& \multicolumn{1}{c}{(\AA)}
& \multicolumn{1}{c}{(\AA)}
& \multicolumn{1}{c}{(\AA)}
& \multicolumn{1}{c}{(cm$^{-2}$)}
& \colhead{ } 
}
\startdata 
\\
\cutinhead{System S1 --- $\langle z_{\rm abs} \rangle$=1.2058}                                                                                
6168.19 & Mg\,{\sc ii}  2796.3543  & 1.2058 &  -90 to 75  & 0.52 & 0.03 & 0.24 & 0.01 & 12.82 $\pm$ 0.04 & \\
6183.93 & Mg\,{\sc ii}  2803.5315  & 1.2058 &  -90 to 75  & 0.32 & 0.03 & 0.14 & 0.01 & 12.82 $\pm$ 0.04 & \\ 
\cutinhead{System S2 --- $\langle z_{\rm abs} \rangle$=1.7876}                                                                               
4315.83 & C\,{\sc iv}   1548.204   & 1.7876 &  -75 to  75 & 0.89 & 0.04 & 0.32 & 0.02 & 14.27 $\pm$ 0.12 & Slightly blended on redward side\\
4322.85 & C\,{\sc iv}   1550.781   & 1.7875 &  -75 to  75 & 0.62 & 0.06 & 0.22 & 0.02 & 14.27 $\pm$ 0.12 & \\
\cutinhead{System S3 --- $\langle z_{\rm abs} \rangle$=2.0138}                                                                                
4200.82 & Si\,{\sc iv}  1393.7602  & 2.0140 &  -75 to  75 & 0.24 & 0.06 & 0.08 & 0.02 & 13.03 $\pm$ 0.16 & \\
4227.58 & Si\,{\sc iv}  1402.7729  & 2.0137 &  -75 to  75 & 0.15 & 0.06 & 0.05 & 0.02 & 13.03 $\pm$ 0.16 & \\ 
4665.94 & C\,{\sc iv}   1548.204   & 2.0138 &  -75 to  75 & 0.57 & 0.04 & 0.19 & 0.01 & 13.77 $\pm$ 0.06 & \\ 
4673.76 & C\,{\sc iv}   1550.781   & 2.0138 &  -75 to  75 & 0.30 & 0.04 & 0.10 & 0.01 & 13.77 $\pm$ 0.06 & \\ 
\cutinhead{System S4 --- $\langle z_{\rm abs} \rangle$=2.1735}                                                                               
4423.51 & Si\,{\sc iv}  1393.7602  & 2.1738 &  -90 to 135 & 1.48 & 0.02 & 0.46 & 0.01 & 13.82 $\pm$ 0.08 \tablenotemark{f} & Upper lim.---blended with N V $\lambda1238$\\ 
4451.76 & Si\,{\sc iv}  1402.7729  & 2.1735 &  -90 to 135 & 0.49 & 0.03 & 0.15 & 0.01 & 13.82 $\pm$ 0.08 \tablenotemark{f} & \\
4844.61 & Si\,{\sc ii}  1526.7070  & 2.1732 &  -90 to 135 & 0.36 & 0.05 & 0.11 & 0.02 & 14.87 $\pm$ 0.20                   & \\ 
4913.56 & C\,{\sc iv}   1548.204   & 2.1737 &  -90 to 135 & 1.09 & 0.06 & 0.34 & 0.02 & 14.48 $\pm$ 0.14 \tablenotemark{f} & \\ 
4921.61 & C\,{\sc iv}   1550.781   & 2.1736 &  -90 to 135 & 0.95 & 0.06 & 0.30 & 0.02 & 14.48 $\pm$ 0.14 \tablenotemark{f} & \\ 
5104.44 & Fe\,{\sc ii}  1608.4511  & 2.1735 &  -90 to 135 & 0.11 & 0.04 & 0.03 & 0.01 & 13.44 $\pm$ 0.09                   & \\ 
5301.81 & Al\,{\sc ii}  1670.7886  & 2.1732 &  -90 to 135 & 0.30 & 0.05 & 0.10 & 0.02 & 13.66 $\pm$ 0.54                   & \\ 
5738.01 & Si\,{\sc ii}  1808.0129  & 2.1737 &  -90 to 135 & 0.13 & 0.03 & 0.04 & 0.01 & 14.87 $\pm$ 0.20                   & \\ 
5885.67 & Al\,{\sc iii} 1854.7184  & 2.1733 &  -90 to 135 & 0.20 & 0.05 & 0.06 & 0.01 & 13.17 $\pm$ 0.19                   & \\  
5911.60 & Al\,{\sc iii} 1862.7910  & 2.1735 &  -90 to 135 & 0.17 & 0.04 & 0.05 & 0.01 & 13.17 $\pm$ 0.19                   & \\ 
7439.05 & Fe\,{\sc ii}  2344.2139  & 2.1734 &  -90 to 135 & 0.36 & 0.05 & 0.11 & 0.02 & 13.44 $\pm$ 0.09                   & \\ 
7535.89 & Fe\,{\sc ii}  2374.4612  & 2.1737 &  -90 to 135 & 0.10 & 0.04 & 0.03 & 0.01 & 13.44 $\pm$ 0.09                   & Noisy\\ 
7561.09 & Fe\,{\sc ii}  2382.7652  & 2.1732 &  -90 to 135 & 0.33 & 0.04 & 0.10 & 0.01 & 13.44 $\pm$ 0.09                   & \\ 
8206.77 & Fe\,{\sc ii}  2586.6500  & 2.1727 &  -90 to 135 & 0.02 & 0.05 & 0.01 & 0.02 & 13.44 $\pm$ 0.09                   & Noisy. In telluric-corrected region\\ 
8250.78 & Fe\,{\sc ii}  2600.1729  & 2.1732 &  -90 to 135 & 0.29 & 0.05 & 0.09 & 0.01 & 13.44 $\pm$ 0.09                   & In telluric-corrected region\\ 
8874.52 & Mg\,{\sc ii}  2796.3543  & 2.1736 &  -90 to 135 & 1.09 & 0.03 & 0.34 & 0.01 & \nodata & Saturated\\ 
8897.26 & Mg\,{\sc ii}  2803.5315  & 2.1736 &  -90 to 135 & 0.96 & 0.03 & 0.30 & 0.01 & \nodata & Saturated\\ 
\cutinhead{System S5 --- $\langle z_{\rm abs} \rangle$=2.1866}
4441.42 & Si\,{\sc iv}  1393.7602  & 2.1866 & -110 to 140 & 0.14 & 0.03 & 0.05 & 0.01 & 12.78 $\pm$ 0.103 & \\ 
\nodata & Si\,{\sc iv}  1402.7729  & \nodata& -110 to 140 & 0.05 & 0.04 & 0.02 & 0.01 & 12.78 $\pm$ 0.103 & Noisy\\
4933.43 & C\,{\sc iv}   1548.204   & 2.1865 & -110 to 140 & 1.00 & 0.06 & 0.31 & 0.02 & 14.03 $\pm$ 0.049 & \\ 
4941.55 & C\,{\sc iv}   1550.781   & 2.1865 & -110 to 140 & 0.53 & 0.06 & 0.17 & 0.02 & 14.03 $\pm$ 0.049 & \\ 
\cutinhead{System S6 --- $\langle z_{\rm abs} \rangle$=2.3156}
4621.09 & Si\,{\sc iv}  1393.7602  & 2.3156 &  -60 to  60 & 0.25 & 0.05 & 0.07 & 0.02 & 12.97 $\pm$ 0.21 & \\ 
5133.30 & C\,{\sc iv}   1548.204   & 2.3156 &  -60 to  60 & 0.23 & 0.03 & 0.07 & 0.01 & 13.33 $\pm$ 0.09 & \\ 
5141.73 & C\,{\sc iv}   1550.781   & 2.3156 &  -60 to  60 & 0.11 & 0.03 & 0.03 & 0.01 & 13.33 $\pm$ 0.09 & \\ 
\cutinhead{System S7 --- $\langle z_{\rm abs} \rangle$=2.3284}                                                                                           
4639.01 & Si\,{\sc iv}  1393.7602  & 2.3284 &  -75 to  85 & 0.33 & 0.05 & 0.10 & 0.02 & 13.04 $\pm$ 0.12 & \\
4668.31 & Si\,{\sc iv}  1402.7729  & 2.3279 &  -75 to  85 & 0.29 & 0.04 & 0.09 & 0.01 & 13.04 $\pm$ 0.12 & Upper lim.---blended with Si II $\lambda1304$\\
5153.13 & C\,{\sc iv}   1548.204   & 2.3285 &  -75 to  85 & 0.32 & 0.03 & 0.10 & 0.01 & 13.43 $\pm$ 0.07 & \\
5161.30 & C\,{\sc iv}   1550.781   & 2.3282 &  -75 to  85 & 0.16 & 0.03 & 0.05 & 0.01 & 13.43 $\pm$ 0.07 & \\
\cutinhead{System S8 --- $\langle z_{\rm abs} \rangle$=2.3801}                                                                               
5233.12 & C\,{\sc iv}   1548.204   & 2.3801 &  -75 to  75 & 0.94 & 0.03 & 0.28 & 0.01 & 14.45 $\pm$ 0.17 &\\ 
5241.89 & C\,{\sc iv}   1550.781   & 2.3802 &  -75 to  75 & 0.82 & 0.04 & 0.24 & 0.01 & 14.45 $\pm$ 0.17 &\\ 
\cutinhead{System S9 --- $\langle z_{\rm abs} \rangle$=2.4376}                                                                               
5321.91 & C\,{\sc iv}   1548.204   & 2.4375 &  -75 to  75 & 0.24 & 0.04 & 0.07 & 0.01 & 13.27 $\pm$ 0.21 & \\ 
5331.09 & C\,{\sc iv}   1550.781   & 2.4377 &  -75 to  75 & 0.05 & 0.04 & 0.02 & 0.01 & 13.27 $\pm$ 0.21 & \\ 
\cutinhead{System S10 --- $\langle z_{\rm abs} \rangle$=2.4649 \phn (Uncertain) }                                                                     
5364.61 & C\,{\sc iv}   1548.204   & 2.4651 &  -55 to  60 & 0.26 & 0.03 & 0.07 & 0.01 & 13.30 $\pm$ 0.10 & Upper lim.---blended with Ni II $\lambda1502$. \\ 
5373.06 & C\,{\sc iv}   1550.781   & 2.4647 &  -55 to  60 & 0.08 & 0.03 & 0.02 & 0.01 & 13.30 $\pm$ 0.10 & \\ 
\cutinhead{System S11 --- $\langle z_{\rm abs} \rangle$=2.4862}                                                                               
4858.82 & Si\,{\sc iv}  1393.7602  & 2.4861 &  -50 to  50 & 0.23 & 0.03 & 0.07 & 0.01 & 13.16 $\pm$ 0.21 & \\ 
4890.39 & Si\,{\sc iv}  1402.7729  & 2.4862 &  -50 to  50 & 0.17 & 0.04 & 0.05 & 0.01 & 13.16 $\pm$ 0.21 & \\ 
5397.33 &  C\,{\sc iv}  1548.204   & 2.4862 &  -50 to  50 & 0.32 & 0.03 & 0.09 & 0.01 & 13.73 $\pm$ 0.23 & \\ 
5406.37 &  C\,{\sc iv}  1550.781   & 2.4862 &  -50 to  50 & 0.23 & 0.03 & 0.07 & 0.01 & 13.73 $\pm$ 0.23 & \\ 
\enddata 
\tablenotetext{a}{Vacuum heliocentric wavelengths.}
\tablenotetext{b}{Velocity range for the equivalent width measurements, relative to $\langle z_{\rm abs} \rangle$ for each system.}
\tablenotetext{c}{Observed-frame equivalent width and 1$\sigma$ random error.}
\tablenotetext{d}{Rest-frame equivalent width and 1$\sigma$ random error.}
\tablenotetext{e}{Ion column density and 1$\sigma$ error computed with VPFIT.}
\tablenotetext{f}{Multiple components fitted with VPFIT at $z_{\rm abs}=2.1734$ and $z_{\rm abs}=2.1741$.}
\tablecomments{The systematic errors are comparable to the random errors.}
\label{tab:intervening_ews}
\end{deluxetable*}
\bigskip

\begin{figure}
\medskip
\centerline{\includegraphics[width=0.70\textwidth]{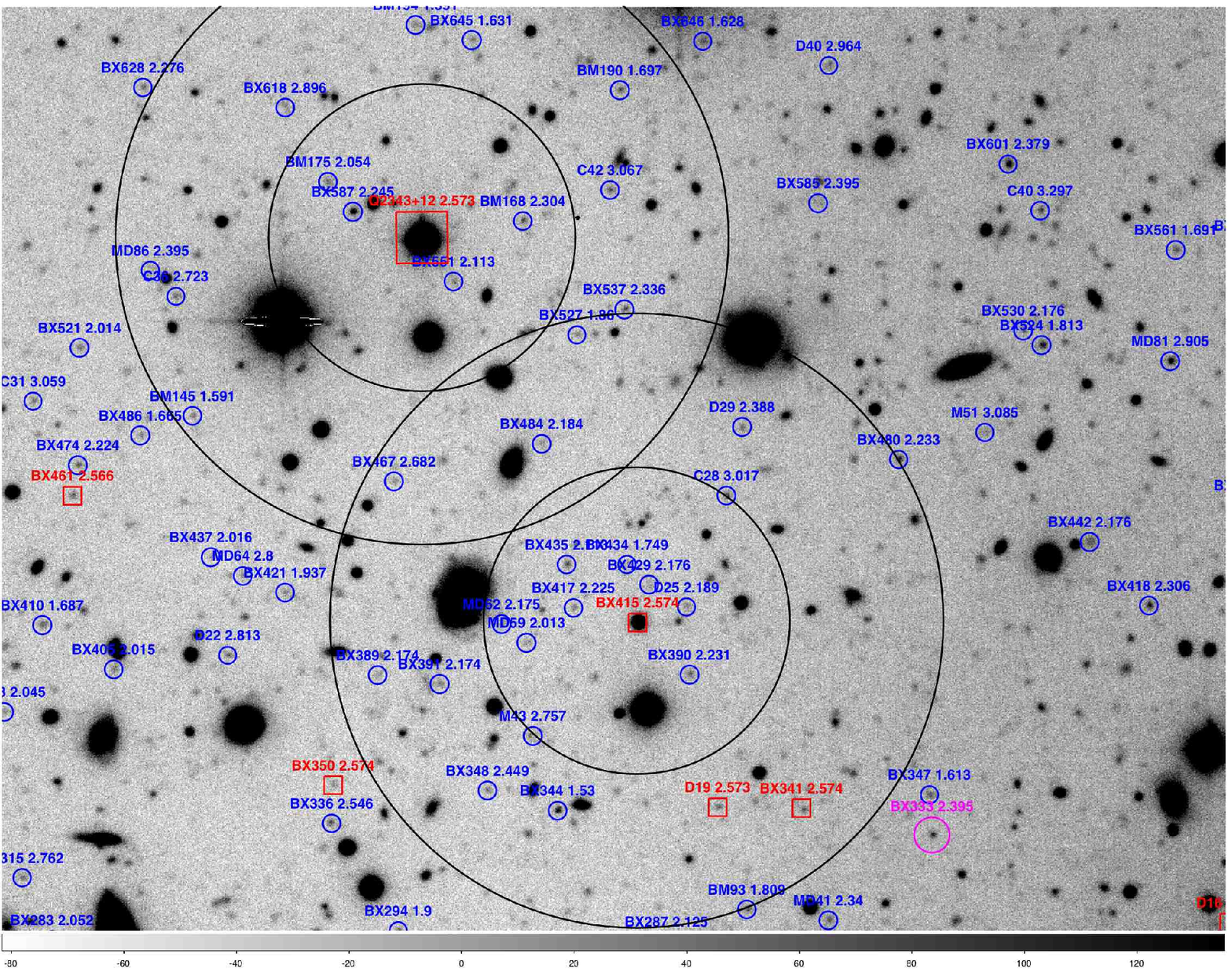}}
\bigskip
\caption{${\cal R}$-band image of the field of the QSOs Q2343+1232 and
Q2343--BX415 obtained with the prime focus camera of the 5.1-m Hale
telescope at Palomar Observatory. The two QSOs 
are at the centres of concentric circles of radii 30 and 
60 arcseconds. At the redshifts of the two quasars,  
$z_{\rm sys} = 2.5727$ and 2.57393 respectively,
the circles correspond to proper transverse distances of 
241 and 481\,kpc respectively. 
Small blue circles indicate galaxies with 
spectroscopically confirmed redshifts; red squares are used
to highlight galaxies at similar redshifts to Q2343+1232
and Q2343--BX415. The galaxy Q2343--BX333, highlighted
with the magenta circle, shows AGN features.\\\label{fig:2343_field}}
\end{figure}

As explained in the Introduction, Q2343--BX415 was discovered during a
survey for galaxies and faint AGN in the field of the brighter QSO
Q2343+1232.  Details of the full survey, which covers $\sim
0.3$\,square degrees on the sky in 13 different fields, are given in
\cite{adelberger05}; Figure~\ref{fig:2343_field} shows the locations
of the two QSOs and of nearby galaxies with measured redshifts.
Within a circle of radius 30\,arcsec centred on Q2343--BX415
(corresponding to a separation of 241~proper~kpc at the redshift of
the quasar), there are 11 galaxies brighter than ${\cal R} = 25.5$
that satisfy the BM and BX color criteria of
\cite{steidel04}\footnote{As discussed by \cite{adelberger04}, the
BM/BX color criteria select galaxies in the redshift interval $z
\simeq 1.5$--2.5\,.}; nine of these have spectroscopically confirmed
redshifts and in all nine cases $z_{\rm gal} \leq z_{\rm QSO}$ (see
Figure~\ref{fig:2343_field}). Similarly, there are 19 BM/BX candidates
at projected distances between 30 and 60\,arcsec; ten have been
confirmed spectroscopically and eight are at redshifts $z_{\rm gal}
\leq z_{\rm QSO}$.

With the survey data in hand,
it may be possible to identify the galaxies producing the 
11 intervening absorption systems in the spectrum of
Q2343--BX415.
In order to assess if any of the 17 galaxies 
at redshifts less or equal to that of the quasar
are associated with any of the 
11 intervening absorption systems, we need to establish:
(a) the systemic redshift, $z_{\rm gal, sys}$,
of candidate absorbing galaxies, and (b) the plausible
redshift difference $\Delta z = |z_{\rm gal, sys} - z_{\rm QSO, abs}|$
between the systemic redshift of a galaxy 
and the absorption redshift of any absorption line system 
that would still allow us to plausibly associate 
the former with the latter.

Point (a) is less straightforward than it may appear at first
sight. While the spectral features of the galaxies can be measured to
within an accuracy of $\sim 50$\kms, more significant uncertainties in
$z_{\rm gal, sys}$ arise due to the existence of significant
velocity fields \emph{within} star-forming galaxies at $z = 2 - 3$, as
evidenced by the several hundred \kms\ difference in the
redshifts measured from different spectral features, such as nebular
emission lines, interstellar absorption lines, and \lya\ emission
\citep[\eg][]{erb04}.  To account for such shifts, we adopted the
following strategy: (i)~For galaxies where the H$\alpha$ emission line
had been detected, we assumed that $z_{\rm sys} = z_{\rm H\alpha}$
with a typical measurement error of $\pm 60$\kms; (ii)~For galaxies
without $z_{\rm H\alpha}$ determinations, we estimated $z_{\rm sys}$
from the average velocity difference between the H$\alpha$ emission
line and the interstellar absorption lines, as measured in a sample of
nearly one hundred BX galaxies at $z \simeq 2.0$--2.6 (Steidel et al,
in preparation): $\langle \Delta v_{{\rm H}\alpha - {\rm IS\,abs}}
\rangle = 164 \pm 130$\kms.

Concerning point (b), we considered a redshift difference  
$\Delta z$ corresponding to $\Delta v  =\pm 300$\,\kms\
sufficient to account for rotation and/or large scale outflows
in the galaxies. Combining points (a) and (b), we then calculated
for each galaxy a redshift range
\begin{equation}
z_{\rm range} = z_{\rm sys} \pm \frac{ (2\delta v + 300) } { c } \, (1 + z_{\rm sys})
\end{equation}
over which it may produce absorption in the spectrum of Q2343--BX415,
where $\delta v$ is the $1 \sigma$ error affecting our determination
of $z_{\rm gal, sys}$ (as discussed above).

\begin{figure*}
\medskip
\begin{center}
\centerline{\includegraphics[angle=270,width=0.8\textwidth]{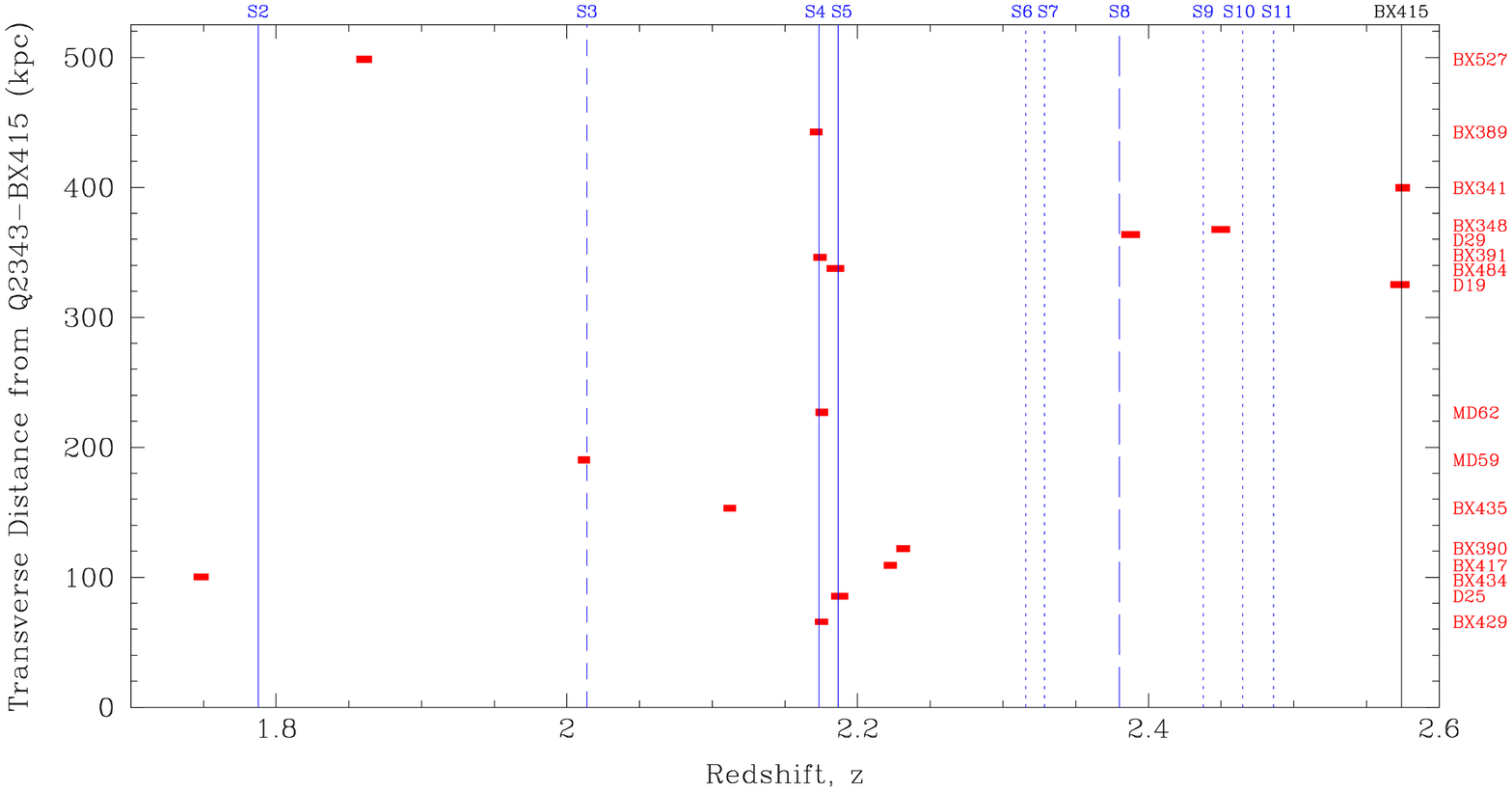}}
\caption{Comparison of the redshifts of ten
intervening absorption systems in line
to Q2343--BX415 with those of 
16 galaxies in the surrounding field.
The absorption redshifts of the
intervening systems are plotted with vertical
lines; the line type gives an indication of the 
strength of the C\,{\sc iv} absorption as follows:
$W_{1548} > 0.3$\,\AA: continuous line; 
$0.2\,$\AA$\,<W_{1548} \leq 0.3$\,\AA: long-dashed;
$0.1\,$\AA$\,<W_{1548} \leq 0.2$\,\AA: short-dashed;
$W_{1548} \leq 0.1$\,\AA: dotted.
The systemic redshift of Q2343--BX415 and its PDLA is also 
shown in a similar way. For comparison, galaxies with spectroscopic
redshifts within $\sim 1$\,arcmin
from Q2343--BX415 are represented by bars,
where the width of each bar denotes the
redshift range over which we estimate that these galaxies 
could produce intervening absorption systems 
(see the text for more details).
The galaxies are identified in the right-hand margin of the panel.\label{fig:intervening_systems}}
\end{center}
\medskip
\end{figure*}

The results of this exercise are summarized in 
Figure~\ref{fig:intervening_systems} 
and Table~\ref{tab:candidate_galaxies}.
In the last column of Table~\ref{tab:candidate_galaxies}
we give our assessment of the likelihood that
a galaxy is the absorber, on a scale from 1 (highest
likelihood) to 3 (lowest likelihood).
Rank 1 is for galaxies that satisfy two conditions:
(i) they are located less than 200~(proper)~kpc from
the line of sight to the quasar, and (ii) the redshift of the absorber is
within the redshift range  
over which the galaxy could produce absorption,
calculated as in equation~(A1). Rank 2 is
for galaxies at impact parameters 
$200 <D /\KPC < 500$ that also satisfy condition (ii),
while rank 3 is assigned to one case where $z_{\rm abs}$ 
is close to, but just outside (by $\sim 125$\,\kms),
$z_{\rm range}$.

Out of the ten intervening systems for which we could potentially find
the absorbing galaxies\footnote{We discount here the Mg\,{\sc ii}
system at $z_{\rm abs} = 1.2058$ (S1) because the BX/BM color criteria
of \cite{steidel04} are inappropriate for finding galaxies at such a
low redshift.}, three of them, S3, S4 and S5, each have one highly
ranked galaxy counterpart which is presumably the absorber. There are
galaxies further away from the line of sight to Q2343--BX415 whose
$z_{\rm range}$ also includes $z_{\rm abs}$ of S4 and S5; in the case
of S4 one of them may be contributing to the absorption, as the line
profiles show more than one component in C\,{\sc iv} (see middle panel
of Figure~\ref{fig:plot_normalised_vpfit.1}).  S8 has a low
probability counterpart, while S2, S6, S7, S9, S10 and S11 do not have
any catalogued galaxies at the right redshifts. This could be due to
the incompleteness of our photometric and spectroscopic catalogs, or
maybe the galaxies responsible for these absorbers are fainter than
${\cal R} = 25.5$.  Perhaps more surprising is the finding that there
are four galaxies at impact parameters $D \lesssim 160$\,kpc,
Q2343--BX434, Q2343--BX417, Q2343--BX390, and Q2343--BX435, that
apparently do not produce absorption lines even to the relatively high
sensitivity limit of our ESI spectrum (corresponding to a $5 \sigma$
equivalent width limit $W_{1548} \simeq 50$\,m\AA).  However, on
closer inspection of the spectrum, we find that certainly in two (and
marginally in three) of these cases, the corresponding C\,{\sc iv}
absorption lines could be blended with other spectral features, which
would make it hard to recognize them, while for Q2343--BX435 we find a
tentative detection of a weak C\,{\sc iv} doublet at $z_{\rm abs} =
2.1096$.

A final comment concerns the fact that 
our as yet incomplete spectroscopic survey 
has already identified four galaxies that are
at redshifts close to that of Q2343--BX415; these
galaxies are indicated by red squares in Figure~\ref{fig:2343_field}.
While their impact
parameters are too large
to associate any of them with the PDLA, they do attest
to the fact that Q2343--BX415 is part of a large scale structure
which includes bright star-forming galaxies as well as the 
bright quasar Q2343+1232.

\newpage
\begin{center}
\begin{deluxetable}{cclcccc}
\tablewidth{0pt}
\tablecaption{\textsc{Candidate Galaxies Associated with the Intervening Absorption Systems}\label{tab:candidate_galaxies}}
\tablehead{
\multicolumn{1}{c}{System} & 
\multicolumn{1}{c}{$\langle z_{\rm abs} \rangle$\tablenotemark{a}} & 
\multicolumn{1}{c}{Candidate} & 
\multicolumn{1}{c}{$z_{\rm gal}$\tablenotemark{b}} &
\multicolumn{1}{c}{${\cal R}$} &
\multicolumn{1}{c}{$D$\tablenotemark{c}} &
\multicolumn{1}{c}{Rank\tablenotemark{d}}\\
\multicolumn{1}{c}{ID} & 
\multicolumn{1}{c}{} &
\multicolumn{1}{c}{Galaxy IDs} & 
\multicolumn{1}{c}{} &
\multicolumn{1}{c}{(mag)} &
\multicolumn{1}{c}{(kpc)} &
\multicolumn{1}{c}{}
}
\startdata 
S1\phn & 1.2058 & \phantom{stri}\nodata & \nodata & \nodata & \nodata & \nodata\\ 
S2\phn & 1.7876 & \phantom{stri}\nodata & \nodata & \nodata & \nodata &\nodata\\ 
S3\phn & 2.0138 & Q2343--MD59            & $2.0116 \pm 0.0042$ & 24.99 & 190 & 1\\ 
S4\phn &\phm{\tablenotemark{e}}2.1735\tablenotemark{e} & Q2343--BX429 & $2.1751 \pm 0.0045$ & 25.12 & \phn66 & 1\\ 
              & & Q2343--MD62            & $2.1752 \pm 0.0045$ & 25.29 & 227 & 2\\ 
              & & Q2343--BX391           & $2.1740 \pm 0.0045$ & 24.51 & 346 & 2\\ 
              & & Q2343--BX389           & $2.1716 \pm 0.0044$ & 24.85 & 443 & 2\\ 
S5\phn & 2.1866 & Q2343--D25             & $2.1877 \pm 0.0060$ & 24.70 & \phn86 & 1\\
              & & Q2343--BX484           & $2.1847 \pm 0.0060$ & 24.77 & 338 & 2\\ 
S6\phn & 2.3156 & \phantom{stri}\nodata & \nodata & \nodata & \nodata & \nodata\\ 
S7\phn & 2.3284 & \phantom{stri}\nodata & \nodata & \nodata & \nodata & \nodata\\ 
S8\phn & 2.3801 & Q2343--D29             & $2.3878 \pm 0.0063$ & 24.39 & 364 & 3\\ 
S9\phn & 2.4376 & \phantom{stri}\nodata & \nodata & \nodata & \nodata & \nodata\\ 
S10    & 2.4649 & \phantom{stri}\nodata & \nodata & \nodata & \nodata & \nodata\\ 
S11    & 2.4862 & \phantom{stri}\nodata & \nodata & \nodata & \nodata & \nodata\\ 
\enddata
\tablenotetext{a}{Mean absorption redshift of the intervening system.}
\tablenotetext{b}{Systemic redshift of the galaxy, 
and the range of redshifts over which we estimate 
that an associated absorbing system could arise (see text for more details).}
\tablenotetext{c}{Impact parameter of candidate galaxy from the line of sight
to Q2343--BX415.}
\tablenotetext{d}{A ranking to indicate the likelihood of the galaxy being associated with the intervening system, where `1' is the highest and `3' the lowest (see text for details).}
\tablenotetext{e}{Multiple components observed in this absorption system.}
\end{deluxetable}
\end{center}

\end{appendix}

\end{document}